\documentclass[11pt, a4paper, sort&compress, numbers, showpacs, preprintnumbers, floatfix]{article}

\usepackage{amsmath}
\usepackage{amsfonts}
\usepackage{amssymb}
\usepackage{graphicx}
\usepackage{bm}

\def\be{\begin{equation}}

\def\ee{\end{equation}}

\def\bi{\bibitem}

\begin{document}

\title{Canonical equivalence, quantization and anisotropic inflation in higher order theory of gravity.}

\author{Subhra Debnath\footnote{E-mail:
subhra\_ dbnth@yahoo.com}~ and Abhik Kumar Sanyal\footnote{E-mail: sanyal\_ ak@yahoo.com}}

\maketitle
\flushbottom

\begin{center}
Dept. of Physics, Jangipur College, Murshidabad, India - 742213
\end{center}

\begin{abstract}
We construct phase-space structure of a typical higher-order theory of gravity, in the background of anisotropic Bianchi-1 mini-superspace, following `Modified Horowitz Formalism' as well as applying `Dirac Algorithm' (after taking care of the divergent terms), and establish equivalence. Canonical quantization, and semiclassical approximation are performed to expatiate the fact that such a quantum theory transits successfully to a classical de-Sitter universe. Inflation has thereafter been studied. The numerical values of the inflationary parameters show excellent agreement with the latest released Planck's data.
\end{abstract}

\section{Introduction}

Inflation in anisotropic models have been studied extensively over decades \cite{1,2,3,4,5,6,7,8,9,10,11}, but higher order theory of gravity has not been considered so far, to the best of our knowledge. It was realized long back that General Theory of Relativity (GTR) is non-renormalizable, and the very early attempt to quantize gravity, required to supplement Einstein-Hilbert action with higher-order curvature invariant terms ($R^2$ and $R_{\mu\nu}R^{\mu\nu}$) \cite{Stelle}. However, perturbative analysis of linearized radiation revealed that there are eight dynamical degrees of freedom, out of which five massive spin-2 excitations are negative definite \cite{Stelle}. Thus the action was supposed to be plagued with ghost degrees of freedom which destroy unitarity of the system \cite{PU}. Nevertheless, there exist convincing  arguments against such perturbative analysis. For example, in the ${1\over N}$ expansion ($N$ being the number of matter fields), such a fourth order gravitational theory was found to be unitary, as $N$ tends to infinity \cite{Tomb77}. Further, the theory was also found to be asymptotically free \footnote{Asymptotically free theories become weak at very short distance, and so are devoid of Landau poles. In asymptotically free theories, the actual physical states have nothing to do with excitations appearing in perturbative theories.} \cite{Tomb80, Frad82, Tomb84}. Additionally, it was also noticed that such excitations are unstable, since they form gauge invariant unitary propagator through S-matrix calculation, and therefore physical S-matrix has to be unitary \cite{Anto86, Tomb15}. Further, ghost were found neither to the leading order under strong coupling expansion in the quantum description of the conformal version of such fourth order gravity theory \cite{Kaku83}, nor in the zero total energy theorem \cite{Boul83}. Nonetheless, it was also claimed as a counter argument that the ghost should appear not only perturbatively but also non-perturbatively \cite{Boul-Deser}. In any case, in the presence of ghost, it is usually argued that a theory is meaningful only below the mass of the ghost, meaning thereby, the ghost mass is the cutoff. This explains why higher curvature gravity theory makes sense as a low energy effective field theory. It is possible that the ghost mass runs during the renormalization flow and the theory admits an ultra-violet fixed point in a non-perturbative analysis. Even then, the ghost mass might turn out to be greater than the scale that we are interested in, and therefore the ghost should be treated perturbatively at a given energy scale. Indeed, there could be other possibilities to deal with a ghost beyond its mass, e.g. involving Fakeons - the fake degrees of freedom, that do not belong to the physical spectrum, but make higher-derivative theories unitary \cite{Anselmi}. We leave the debate and mention that later, with the advent of superstring, heterotic string and supergravity theories, it is now quite evident that under weak field approximation, all these theories primarily reduce to effective actions containing higher-order curvature invariant terms in $4$-dimensions \cite{SUSY, SUGRA}, and we feel it worth studying the evolution of early universe in view of such higher-order theories. The reason being, even if ghosts exist, it can sometimes be dealt with at least in minisuperspace model. If not, particularly for the action under consideration (say), the technique we develop here may be applied to ghost-free theories, for example, in the bigravity unitary theories free from the Boulware-Deser ghost \cite{Paulos} in future, as it has already been applied in modified Lanczos-Lovelock gravity, although in isotropic model \cite{Ab00, Ab0}.\\

Turning our attention to the present issue, it should be mentioned that although, inflation occurred at the sub-Planckian epoch, sometime between ($10^{-36} - 10^{-26}$)~sec., curvature was too strong at that epoch, and it is not wise to make further weak field approximation and reduce the action to GTR. This fact motivates us to study inflation in anisotropic model, with higher-order curvature invariant terms. Inflation is essentially a quantum theory of perturbation, and as mentioned, has occurred in the sub-Planckian era. Since a complete quantum theory of gravity is still not at hand, it is reasonable to explore some physical insight near Planck's era, through `quantum cosmology'. It is therefore necessary to understand how a quantum universe evolves through a semiclassical domain to give a pathway to the inflationary regime, and thereafter how inflation halts and ends up with a matter dominated era, dubbed as graceful exit. This was attempted earlier in the isotropic cosmological model, with different fourth order gravity theories. It was found to render a hermitian Hamiltonian \cite{Ab1, Ab2, Ab3, Ab4, Ab5, Ab6, Ab7, Ab8, Ab9}, with a well-behaved quantum dynamics followed by an inflationary era. In the present manuscript, we extend our work in the background of an anisotropic Bianchi-1 mini-superspace model.\\

To build a quantum universe with higher-order curvature invariant terms, canonical formulation is necessitated, which is a non-trivial task. This is only possible by seeking additional degrees of freedom, which was initiated by Ostrogradski long back \cite{Os}. However, if the Hessian determinant vanishes resulting in a singular Lagrangian, the formalism \cite{Os} does not apply, and Dirac's constrained analysis \cite{D1, D2} is invoked. For a positive definite fourth order gravitational action, Horowitz' proposed a technique that bypasses Dirac's constraint analysis elegantly even if the Hessian determinant ceases to exist \cite{Horo}. The technique may as well be applied in general for higher order theories other than gravity. It is important to mention that Dirac's constrained analysis and Horowitz' formalism yield identical phase-space structure \cite{Ab3}. Nevertheless, since Horowitz' technique is plagued with some serious problems \cite{Pol, Ab1, Ab2}, a modified Horowitz' formalism (MHF) was developed \cite{Ab1, Ab2, Ab3, Ab4, Ab5, Ab6, Ab7, Ab8, Ab9}. Let us discuss the essential features of MHF in brief. While canonical (ADM) formulation of GTR is possible only in terms of the basic variables $h_{ij}$ - the induced three-metric \cite{ADM}, the same for gravitational theories with higher-order curvature invariant terms, requires additional degrees of freedom, as mentioned. The extrinsic curvature tensor - $K_{ij}$ plays the role of these additional basic variables \footnote{It is important to mention that the classical Hamiltonian may be formulated with arbitrary set of variables which are canonically related. However such canonical transformations cannot be translated automatically after quantization, particularly in the higher-order theories, due to non-linearities. Thus, quantum equations with different set of variables lead to different results. A viable quantum theory of gravity requires at least a hermitian effective Hamiltonian that allows an appropriate semiclassical approximation, so that finally one ends up with the universe we live in.}. The main essence of MHF formalism is primarily to express the action in terms of the basic variables $h_{ij}$, and remove all the available divergent terms, upon integration by parts. Thereafter, auxiliary variables, found by taking the derivatives of the `action' with respect to the highest derivatives appearing in it, are judiciously plugged into the action, so that the action turns out to be canonical. The Hamiltonian is then expressed with respect to the basic variables and the auxiliary variables, establishing diffeomorphic invariance. However, the Hamiltonian so obtained is not well suited for quantization, since it contains momenta - canonically conjugate to the auxiliary variables, with higher than second degree. Hence, finally a well behaved Hamiltonian is constructed under the replacement of the auxiliary variables by the basic variables $K_{ij}$ - the extrinsic curvature tensor, following canonical transformations \footnote{This fact again reveals that not all canonically related variables are suitable for quantization.}. In the process, one ends up with a viable phase-space structure of the Hamiltonian. The Hamiltonian so obtained (so far, in the isotropic background) is found to be well behaved in every respect.\\

As, already mentioned, Dirac's and Horowitz' formalisms produce identical Hamiltonian, and therefore plagued with the same disease. This is beyond contemplation, due to extensive mathematical rigour of Dirac's formalism. Later, it was noticed that Dirac's formalism after taking care of the divergent terms appearing in the action, leads to Hamiltonian, identical to the one obtained following MHF \cite{Ab9, Ab10, Ab11}. This implies that the divergent terms indeed play a crucial role in higher order theories. Such equivalence has been established earlier, only in the isotropic model. Very recently, it has been proved in a class of anisotropic models too \cite{Ani}. Although Hamiltonian obtained following other techniques are canonically related to the one obtained following MHF, such transformation can not be extended in the quantum domain due to nonlinearity, as already mentioned. Thus, the important observation is: although in the classical domain, total derivative terms do not play any significant role, they indeed play a vital role in the quantum domain \cite{Ab9}.\\

In the following section, we take up Bianchi-1 metric, write down the the field equations and corresponding de-Sitter solutions in vacuum for an action presented by Stelle \cite{Stelle}, but with functional coupling parameters. In section 3, we proceed to construct the phase-space structure of the Hamiltonian following Dirac's constrained analysis (after taking care of the divergent terms). For completeness, we present MHF in appendix A. Canonical quantization and semiclassical approximation have been performed in section 4. In section 5, we study inflation and conclude in section 6.

\section{Action, Field equations and Classical de-Sitter solution:}

Our starting point is the following general fourth-order gravitational action,

\be\label{A} A=\int\Bigg[\alpha(\phi)R + \beta(\phi)R^2 + \gamma(\phi)R_{\mu\nu}R^{\mu\nu} - \frac{1}{2}\phi_{,\mu}\phi^{,\mu} - V(\phi) \Bigg]\sqrt{-g}d^4x, \ee
in which all the coupling parameters are arbitrary functions of the scalar field $\phi$. Varying the above action \eqref{A} with respect to the metric, following field equation is found:

\be\begin{split} \alpha G_{\mu\nu} &+ g_{\mu\nu}\Box\alpha - \alpha_{;\mu\nu} + \beta\Big(2RR_{\mu\nu} - \frac{1}{2}g_{\mu\nu}R^2\Big)  + 2g_{\mu\nu}\Box (\beta R) - 2(\beta R)_{;{\mu\nu}}\\
& + \gamma\Big(2R^{\alpha\beta}R_{\alpha\mu\beta\nu} - \frac{1}{2}g_{\mu\nu}R_{\alpha\beta}R^{\alpha\beta}\Big) + \frac{1}{2}g_{\mu\nu}\Box (\gamma R) + \Box (\gamma R_{\mu\nu}) - (\gamma R)_{;{\mu\nu}} = T_{\mu\nu},\end{split}\ee
where, $G_{\mu\nu} = R_{\mu\nu} - \frac{1}{2}g_{\mu\nu}R$ is the Einstein tensor, while $T_{\mu\nu} = \nabla_\mu\phi\nabla_\nu\phi - \frac{1}{2}g_{\mu\nu}\nabla_\lambda\phi\nabla^\lambda\phi - g_{\mu\nu}V(\phi)$ is the energy-momentum tensor. Variation of the action \eqref{A} with respect to $\phi$ yields the following generalized Klein-Gordon equation,

\be \Box\phi - \alpha'R - \beta' R^2 - \gamma'R_{\mu\nu}R^{\mu\nu} - V' = 0,\ee
where prime denotes derivative with respect to $\phi$. Let us now express the homogeneous, anisotropic, and axially symmetric Bianchi-I minisuperspace model in the following form:

\be\label{aniso}\begin{split} & ds^2 = - N(t)^2dt^2 + A(t)^2 dx^2 + B(t)^2(dy^2 + dz^2).\end{split}\ee
Correspondingly, the Ricci scalar and the square of the Ricci tensor are given by,

\be \begin{split} &\hspace{1.5 in}R = \frac{2}{N^2} \Bigg(\frac{\ddot A}{A} + 2 \frac{\ddot B}{B} + 2\frac{\dot A}{A}\frac{\dot B}{B} + \frac{\dot B^2}{B^2} - \frac{\dot A}{A}\frac{\dot N}{N} - 2\frac{\dot B}{B}\frac{\dot N}{N}\Bigg),\\&
R_{\mu\nu}R^{\mu\nu} = \frac{1}{N^4} \Bigg[\left(\frac{\ddot A}{A} + 2\frac{\ddot B}{B} - \frac{\dot A}{A}\frac{\dot N}{N} - 2\frac{\dot B}{B}\frac{\dot N}{N}\right)^2 + \left(\frac{\ddot A}{A} + 2 \frac{\dot A}{A}\frac{\dot B}{B} - \frac{\dot A}{A}\frac{\dot N}{N}\right)^2 + 2\left(\frac{\ddot B}{B} + \frac{\dot A}{A}\frac{\dot B}{B} + \frac{\dot B^2}{B^2}  - \frac{\dot B}{B}\frac{\dot N}{N}\right)^2\Bigg].\end{split}\ee
The ($^0_0$) equation of Einstein, the $\phi$ variation equation, together with the expressions for the expansion ($\theta$) and the shear ($\sigma$) scalars respectively are the following,

\be\label{00}\begin{split} &2\alpha\Bigg(\frac{\dot B^2}{B^2} + 2\frac{\dot A\dot B}{AB} \Bigg) + 4\beta\Bigg(2\frac{\dot A\dddot A}{A^2} + 4\frac{\dot B\dddot A}{AB} + 4\frac{\dot A\dddot B}{AB} + 8\frac{\dot B\dddot B}{B^2} - \frac{\ddot A^2}{A^2} - 2\frac{\dot A^2\ddot A}{A^3} + 8\frac{\dot B^2\ddot A}{AB^2} - 4\frac{\ddot B^2}{B^2} + 4\frac{\dot A^2\ddot B}{A^2B}\\
& + 8\frac{\dot A\dot B\ddot B}{AB^2} - 4\frac{\ddot A\ddot B}{AB} - 4\frac{\dot A^3\dot B}{A^3B} - 8\frac{\dot A^2\dot B^2}{A^2B^2} - 8\frac{\dot A\dot B^3}{AB^3} - 7\frac{\dot B^4}{B^4} \Bigg) + 2\gamma\Bigg(2\frac{\dot A\dddot A}{A^2}+ 2\frac{\dot B\dddot A}{AB} + 2\frac{\dot A\dddot B}{A B} + 6\frac{\dot B\dddot B}{B^2}\\
& - \frac{\ddot A^2}{A^2} - 2\frac{\dot A^2\ddot A}{A^3} + 2\frac{\dot A\dot B\ddot A}{A^2B} - 3\frac{\ddot B^2}{B^2} + 2\frac{\dot A^2\ddot B}{A^2B} + 6\frac{\dot A\dot B\ddot B}{AB^2} - 2\frac{\ddot A\ddot B}{AB} - 2\frac{\dot A^3\dot B}{A^3B} - 7\frac{\dot A^2\dot B^2}{A^2B^2} - 4\frac{\dot A\dot B^3}{AB^3} - 5\frac{\dot B^4}{B^4} \Bigg)\\
& + 2\alpha'\dot\phi\Bigg(\frac{\dot A}{A} + 2\frac{\dot B}{B}\Bigg) + 8\beta'\dot\phi\Bigg(\frac{\dot A\ddot A}{A^2} + 2\frac{\dot B\ddot A}{AB} + 2\frac{\dot A\ddot B}{AB} + 4\frac{\dot B\ddot B}{B^2} + 2\frac{\dot A^2\dot B}{A^2B} + 5\frac{\dot A\dot B^2}{AB^2} + 2\frac{\dot B^3}{B^3}\Bigg)\\& + 4\gamma'\dot\phi\Bigg(\frac{\dot A\ddot A}{A^2} + \frac{\dot B\ddot A}{AB} + \frac{\dot A\ddot B}{AB} + 3\frac{\dot B\ddot B}{B^2}  + \frac{\dot A^2\dot B}{A^2B} + \frac{\dot A\dot B^2}{AB^2} + \frac{\dot B^3}{B^3}\Bigg) - \Bigg(\frac{\dot\phi^2}{2} + V\Bigg)= 0,\end{split}\ee

\be\label{phi}\begin{split} &\ddot\phi + \Bigg(\frac{\dot A}{A} + 2\frac{\dot B}{B}\Bigg)\dot\phi + V' - 2\alpha'\Bigg(\frac{\ddot A}{A} + 2\frac{\ddot B}{B} + 2\frac{\dot A\dot B}{A B} + \frac{\dot B^2}{B^2}\Bigg)\\& - 4\beta'\Bigg(\frac{\ddot A^2}{A^2} + 4\frac{\dot A\dot B\ddot A}{A^2B} + 2\frac{\dot B^2\ddot A}{AB^2} + 4\frac{\ddot B^2}{B^2} + 8\frac{\dot A\dot B\ddot B}{AB^2} + 4\frac{\dot B^2\ddot B}{B^3}  + 4\frac{\ddot A\ddot B}{AB} + 4\frac{\dot A^2\dot B^2}{A^2B^2} + 4\frac{\dot A\dot B^3}{AB^3} + \frac{\dot B^4}{B^4} \Bigg) \\
&- 2\gamma'\Bigg(\frac{\ddot A^2}{A^2} + 2\frac{\dot A\dot B\ddot A}{A^2B} + 3\frac{\ddot B^2}{B^2} + 2\frac{\dot A\dot B\ddot B}{AB^2} + 2\frac{\dot B^2\ddot B}{B^3} + 2\frac{\ddot A\ddot B}{AB}+ 3\frac{\dot A^2\dot B^2}{A^2B^2} + 2\frac{\dot A\dot B^3}{AB^3} + \frac{\dot B^4}{B^4}\Bigg) = 0,\end{split}\ee

\be \label{theta}\begin{split}&\theta = {v^{\mu}}_{;\mu} = {\dot A\over A} + 2{\dot B\over B};\hspace{0.4 cm} \sigma^2 = {1\over 2}\sigma_{\mu\nu}\sigma^{\mu\nu}= {1\over 3}\left[{\dot A\over A} - {\dot B\over B}\right]^2,\\&
\mathrm{where,}~~\sigma_{\mu\nu} = v_{(\mu;\nu)} + {1\over 2}(v_{\mu;\alpha}v^{\alpha}v_{\nu}+v_{\nu;\alpha}v^{\alpha}v_{\mu}) -{1\over \theta}(g_{\mu\nu} + v_\mu v_\nu).\end{split}\ee
In the above, $v_\mu$ represents the four-velocity, so that $v_\mu v^\mu = -1$. The field equations, (\ref{00}) and (\ref{phi}) admit the following de-Sitter solutions:
\be\label{Sol1}\begin{split}& \mathrm{\bf{Set-I}}\\
& A = A_0 e^{4\lambda t},~~B = B_0 e^{\lambda t},  ~\phi = \phi_0 e^{-\lambda t},\\&
\mathrm{along~with}, ~V = V_0\phi^2, ~\alpha = \alpha_0\phi^2, ~\beta = \beta_0\phi^2,  ~\gamma = \gamma_0\phi^2,~\\&
\mathrm{where},~\alpha_0 = \frac{1}{27}\left(\frac{V_0}{\lambda^2} - \frac{23}{20}\right),~ \gamma_0 = - \left(\frac{1}{4860\lambda^2} + \frac{V_0}{972\lambda^4} + 3\beta_0\right),\\&
\mathrm{so~that,}~~
AB^2=A_0B_0^2 e^{6\lambda t}, ~~a(t) = (A_0B_0^2)^{1\over 3}e^{2\lambda t},~~\theta = 6\lambda;~~\sigma^2 = 3\lambda^2.\end{split}\ee

\be\label{Sol2}\begin{split}& \mathrm{\bf{Set-II}}\\
& A = A_0 e^{\frac{\sqrt 3 + 2c}{\sqrt 3}\lambda t},~~B = B_0 e^{\frac{\sqrt 3 - c}{\sqrt 3}\lambda t}, ~\phi = \phi_0 e^{-\lambda t},\\&
\mathrm{together~with},~V = V_0\phi^2, ~\alpha = \alpha_0\phi^2, ~\beta = \beta_0\phi^2, ~\gamma = \gamma_0\phi^2, \\&
\mathrm{where},~\alpha_0 = \frac{1}{c^2 + 6}\left(\frac{V_0}{\lambda^2} - \frac{5c^2 + 18}{4(2c^2 + 9)}\right),\\&
\gamma_0 = \frac{1}{2c^2 + 3}\left[\frac{c^2}{4(c^2 + 6)(2c^2 + 9)\lambda^2} - \frac{V_0}{2(c^2 + 6)\lambda^4} - 2\beta_0(c^2 + 6)\right],\\&
\mathrm{so~that,}~~
AB^2 = A_0B_0^2 e^{3\lambda t},~a(t) =(A_0B_0^2)^{1\over 3}e^{\lambda t},~~\theta = 3\lambda,~~\sigma^2 = c^2\lambda^2.\end{split}\ee
In the above, $AB^2$ is the three volume, $a(t) = (AB^2)^{1\over 3}$ is the average expansion scale factor, $\theta$ and $\sigma$ are the expansion and the shear scalar respectively, while, $A_0, B_0, \phi_0, V_0, \alpha_0, \beta_0, \gamma_0, \lambda, c$ are constants. It is interesting to note that in both the sets of de-Sitter solutions presented above, only two out of $\alpha_0,~\beta_0,~\gamma_0,~V_0$ remain arbitrary. Further, although the expansion scalar and shear scalar are different in the two sets (unless $c^2 = 3$ in the second set, which keeps $B(t)$ constant and the universe expands unidirectionally), the forms of all the parameters of the theory ($\alpha, \beta, \gamma$) including the potential $V(\phi)$ remain unaltered.

\section{Canonical formulation:}

As already mentioned in the introduction, the Modified Horowitz' Formalism (MHF) and Dirac's algorithm (after controlling all the divergent terms appearing in the action) lead to identical phase-space structure of the Hamiltonian, in isotropic and homogeneous space-time, as well as in a class of anisotropic models \cite{Ani}. In this section, we construct the phase-space structure of the Hamiltonian corresponding to the action \eqref{A} in the anisotropic minisuperspace model under consideration \eqref{aniso} applying Dirac's algorithm. To prove the fact that we have produced correct Hamiltonian, MHF has been performed in appendix A. For convenience (the reason will transpire later), we initiate our programme with the following form of anisotropic Bianchi-1 space-time,

\be\label{anisonew} ds^2 = - N(t)^2dt^2 + e^{2\xi}\Big[e^{-4\chi} dx^2 + e^{2\chi}\big(dy^2 + dz^2\big)\Big],\ee
in view of the transformations: $A = e^{(\xi - 2\chi)}$ and $B = e^{(\xi + \chi)}$. As a result, the average (isotropic) scale factor is given by $a(t) = (AB^2)^{1\over 3} = e^\xi$, and the anisotropy is characterized by $\big(\frac{B}{A}\big)^{1\over3} = e^\chi$. Further, under the above transformation, the expressions for the expansion scalar is $\theta = 3\dot \xi$ and that for the shear scalar is $\sigma^2 = 3\dot\chi^2$, in view of \eqref{theta}. Hence, $H = \dot\xi$ represents the isotropic Hubble expansion rate, while $\Sigma = \dot\chi$ measures the anisotropic expansion rate. We repeat: although classically one can formulate the phase-space structure starting from different set of variables and can find Hamiltonians related under canonical transformation; quantum mechanically, such transformations do not work, in the non-linear theories, in particular. It is therefore suggestive to cast the Hamiltonian in terms of the basic variables, $\{{h_{ij}, K_{ij}}\}$, where $h_{ij}$ is the induced three-metric, and $K_{ij}$ is the extrinsic curvature tensor. In order to remove true divergent terms from the action (obtainable under variational principle), both in Dirac's formalism and MHF, one should start with basic variables, $h_{11} = z\delta_{11} = e^{2\xi}\delta_{11}$ and $h_{22} = h_{33} = y\delta_{jj} = e^{2\chi}\delta_{jj}$ ($j = 2, 3$) \footnote{It was noticed earlier \cite{Ab1, Ab2} that otherwise, some additional divergent terms are removed, which do not appear from variational principle.}. In view of the basic variables $h_{ij}$, the action (\ref{A}) is expressed as,

\be\label{A1}\begin{split}
A = &\int\Bigg[\frac{\alpha(\phi)}{N}\Bigg(3\frac{\ddot z}{z} - 3\frac{\dot N\dot z}{Nz} + \frac{3\dot y^2}{2y^2}\Bigg) + \frac{\beta(\phi)}{N^3}\Bigg(9\frac{\ddot z^2}{z^2} - 18\frac{\dot N\dot z\ddot z}{Nz^2} + 9\frac{\dot y^2\ddot z}{y^2z}  + 9\frac{\dot N^2\dot z^2}{N^2z^2} - 9\frac{\dot N\dot y^2\dot z}{Ny^2z} + \frac{9\dot y^4}{4y^4}\Bigg)\\
& + \frac{\gamma(\phi)}{N^3}\Bigg(3\frac{\ddot z^2}{z^2} - \frac{3\dot z^2 \ddot z}{2z^3} - 6\frac{\dot N\dot z\ddot z}{Nz^2} + \frac{9\dot y^2\ddot z}{2y^2z}+ \frac{3\ddot y^2}{2y^2} + \frac{9\dot y\dot z\ddot y}{2y^2z} - 3\frac{\dot y^2\ddot y}{y^3} - 3\frac{\dot N\dot y\ddot y}{Ny^2} + \frac{3\dot z^4}{4z^4} + \frac{3\dot N\dot z^3}{2Nz^3}\\
& + \frac{9\dot y^2 \dot z^2}{8y^2 z^2} + 3\frac{\dot N^2\dot z^2}{N^2z^2} - \frac{9\dot y^3\dot z}{2y^3z} - 9\frac{\dot N\dot y^2\dot z}{Ny^2z} + \frac{15\dot y^4}{4y^4} + 3\frac{\dot N\dot y^3}{Ny^3} + \frac{3\dot N^2\dot y^2}{2N^2y^2}\Bigg) + \Bigg(\frac{\dot\phi^2}{2N} - NV(\phi)\Bigg) \Bigg]z^{3\over2}dt.\end{split}\ee
The primary observation is: despite being a Lagrange multiplier, the lapse function $N(t)$ appears in the action with its time derivative, unlike GTR. This uncanny behaviour which insists to treat the lapse function as a dynamical variable, restrains from establishing diffeomorphic invariance. However, one can easily compute the Hessian determinant to be sure that it vanishes, making the action degenerate. Thus, canonical formulation requires to handle the situation following Dirac's algorithm of constraint analysis. To proceed further, let us first integrate the action (\ref{A1}) by parts as already mentioned, and remove following total derivative terms,

\be\label{Sigma} \Sigma = \Bigg[3\frac{\alpha(\phi)}{N}\left(\frac{\dot z}{z}\right) - \frac{\gamma(\phi)}{N^3}\Bigg(\frac{\dot z^3}{2z^3} + \frac{\dot y^3}{y^3}\Bigg)\Bigg]z^{3\over2}.\ee
We mention that in the case of Dirac's formalism, since $h_{ij}$ and $K_{ij}$ are kept fixed at the boundary, these total derivative terms \eqref{Sigma} trivially vanish. The action (\ref{A1}) now reads as,

\be\label{A2}\begin{split}
A =& \int\Bigg[\frac{\alpha}{N}\Bigg(\frac{3\dot y^2}{2y^2} - \frac{3\dot z^2}{2z^2}\Bigg) - 3\frac{\alpha'\dot\phi}{N}\frac{\dot z}{z} + \frac{\beta}{N^3}\Bigg(9\frac{\ddot z^2}{z^2} - 18\frac{\dot N\dot z\ddot z}{Nz^2} - 9\frac{\dot y^2\ddot z}{y^2z} + 9\frac{\dot N^2\dot z^2}{N^2z^2} + \frac{9\dot y^4}{4y^4} - 9\frac{\dot N\dot y^2\dot z}{Ny^2z} \Bigg)\\& + \frac{\gamma}{N^3}\Bigg(3\frac{\ddot z^2}{z^2} - 6\frac{\dot N\dot z\ddot z}{Nz^2} + \frac{9\dot y^2\ddot z}{2y^2z} + \frac{3\ddot y^2}{2y^2} + \frac{9\dot y\dot z\ddot y}{2y^2z} - 3\frac{\dot N\dot y\ddot y}{Ny^2} + \frac{9\dot y^2 \dot z^2}{8y^2 z^2} + 3\frac{\dot N^2\dot z^2}{N^2z^2} - 3\frac{\dot y^3\dot z}{y^3z} - 9\frac{\dot N\dot y^2\dot z}{Ny^2z}\\&
+ \frac{3\dot y^4}{4y^4} + \frac{3\dot N^2\dot y^2}{2N^2y^2}\Bigg) + \frac{\gamma'\dot\phi}{N^3}\Bigg(\frac{\dot z^3}{2z^3} + \frac{\dot y^3}{y^3}\Bigg) + \Bigg(\frac{\dot\phi^2}{2N} - NV\Bigg)\Bigg]z^{3\over2}dt. \end{split}\ee
The above action \eqref{A2} being free from all the divergent terms (Note that further integration by parts only introduces higher derivative terms), is our starting point for both the MHF and Dirac's constrained analysis.

\subsection{Dirac formalism:}

In order to study the phase-space structure of action (\ref{A}) for the anisotropic Bianchi-1 space-time \eqref{aniso} following Dirac’s algorithm, let us make change of variables, $x = \frac{\dot z}{N}$ and $w = \frac{\dot y}{N}$ in the action (\ref{A2}). Further, treating $(\frac{\dot z}{N} - x)$ and $(\frac{\dot y}{N} - w)$ as constraints, we insert these terms through Lagrange multipliers $\lambda$ and $\tau$ in the associated point Lagrangian to obtain,

\be\label{L}\begin{split} L = &\Bigg[\frac{N\alpha}{2}\Bigg( \frac{w^2}{y^2} - \frac{x^2}{z^2}\Bigg) - \alpha'\dot\phi\frac{x}{z} + 9\beta\Bigg(\frac{\dot x^2}{Nz^2} + \frac{w^2\dot x}{y^2z} - \frac{Nw^4}{4y^4} \Bigg) \\& + 3\gamma\Bigg(\frac{\dot x^2}{Nz^2} + \frac{3w^2\dot x}{2y^2z} + \frac{\dot w^2}{2Ny^2} + \frac{3wx\dot w}{2y^2z} + \frac{Nw^4}{4y^4} - \frac{Nw^3x}{y^3z}
+ \frac{3Nw^2x^2}{8y^2z^2} \Bigg)\\& + \gamma'\dot\phi\Bigg(\frac{x^3}{2z^3}  + \frac{w^3}{y^3}\Bigg)
+ \Bigg(\frac{\dot\phi^2}{2N} - NV\Bigg)\Bigg]z^{3\over2} + \lambda\Bigg(\frac{\dot z}{N} - x\Bigg) + \tau\Bigg(\frac{\dot y}{N} - w\Bigg).\end{split}\ee
One can clearly observe that the above point Lagrangian \eqref{L} is now cured from the disease of having time derivative of lapse function $N$. The corresponding momenta are,

\be\label{momenta}\begin{split}& p_x = 3\Bigg[3\beta\Bigg(\frac{2\dot x}{Nz} + \frac{w^2}{y^2}\Bigg) + \gamma\Bigg(\frac{2\dot x}{Nz} + \frac{3w^2}{2y^2}\Bigg)\Bigg]\sqrt z\\
& p_w = 3\gamma\Bigg(\frac{\dot w}{N} + \frac{3wx}{2z}\Bigg)\frac{z^{3\over2}}{y^2};~~~~p_z = \frac{\lambda}{N};~~~~p_y = \frac{\tau}{N};~~~~p_\tau = p_N=0=p_\lambda\\
& p_\phi = \Bigg[- 3\alpha'\frac{x}{z} + \gamma'\Bigg(\frac{x^3}{2z^3}  + \frac{w^3}{y^3}\Bigg) + \frac{\dot\phi}{N}\Bigg]z^{3\over2}\end{split}\ee
The Hamiltonian constraint therefore reads as,

\be\label{H_c}\begin{split} H_c &= \dot zp_z + \dot xp_x + \dot yp_y + \dot wp_w + \dot \phi p_\phi + \dot Np_N + \dot\lambda p_\lambda + \dot\tau p_\tau - L\\
&=\Bigg[\frac{3\alpha}{2}\left(\frac{x^2}{z^2} + \frac{w^2}{y^2}\right)
+ 9\beta\Bigg(\frac{\dot x^2}{N^2z^2} - \frac{w^4}{4y^4}\Bigg)
+ 3\gamma\Bigg(\frac{\dot x^2}{N^2z^2} + \frac{\dot w^2}{2N^2y^2} - \frac{w^4}{4y^4} + \frac{w^3x}{y^3z} - \frac{3w^2x^2}{8y^2z^2}\Bigg)\\
&~~~~~~ + \left(\frac{\dot\phi^2}{2N^2} + V\right)\Bigg]Nz^{3\over2} + \lambda x + \tau w.\end{split}\ee
Now, from the expressions of momenta (\ref{momenta}) we find,

\be\label{dot}\begin{split} &\dot x = \frac{N}{2(3\beta + \gamma)}\Bigg[\frac{\sqrt{z}p_x}{3} - (2\beta + \gamma)\frac{3w^2z}{2y^2}\Bigg];\hspace{0.75cm}\dot w = \frac{N}{z}\Bigg[\frac{y^2p_w}{3\sqrt z\gamma} - \frac{3}{2}wx\Bigg];\hspace{0.75cm}
\dot\phi = N\Bigg[\frac{p_\phi}{z^{3\over2}} + U\Bigg],\end{split}\ee
where, $U(x,y,z,w,\phi)=3\alpha'\frac{x}{z} - \gamma'\Big(\frac{x^3}{2z^3} + \frac{w^3}{y^3}\Big).$ Using these expressions (\ref{dot}), it is now possible to express the Hamiltonian (\ref{H_c}) as,

\be\label{H_c1}\begin{split} H_c = &\lambda x + \tau w + N\Bigg[\frac{\sqrt zp_x^2}{12(3\beta + \gamma)} + \frac{y^2p_w^2}{6\gamma z^{3\over2}} - \frac{3(2\beta + \gamma)w^2zp_x}{4(3\beta + \gamma)y^2} - \frac{3wxp_w}{2z} + \frac{p_\phi^2}{2z^{3\over2}} + U p_\phi\\& + \Bigg\{\frac{U^2}{2}
 + \frac{3\gamma(12\beta + 5\gamma)w^4}{16(3\beta + \gamma)y^4} + \frac{3\alpha}{2}\Bigg(\frac{x^2}{z^2} - \frac{w^2}{y^2}\Bigg)
+ 3\gamma\Bigg(\frac{w^3x}{y^3z} + \frac{3w^2x^2}{4y^2z^2}\Bigg) + V\Bigg\}z^{3\over2}\Bigg].\end{split}\ee
The definition of momenta \eqref{momenta} reveals that we require four primary constraints involving Lagrange multipliers or their conjugates viz,

\be\label{constraints}\phi_1 = Np_z - \lambda \approx 0,~ \phi_2 = p_\lambda \approx 0,~\phi_3 = Np_y - \tau \approx 0,~ \phi_4 = p_\tau \approx 0.\ee
Note that the constraint $\phi_5 = p_N$ associated with lapse function $N$ vanishes strongly, since it is non-dynamical, and may therefore be safely ignored. The above four primary constraints \eqref{constraints} are second class, since, they have non-vanishing Poisson bracket with  other constraints. There are two ways to handle the second-class constraints. 1. The Hamiltonian may be extended by adding the constraints with arbitrary coefficients $u$’s to it, and then may be solved for the consistency equations which yields $u$’s unambiguously due to the fact that $det |{\phi_i,\phi_j}| \neq 0$. 2. Dirac bracket may be introduced and the the constraints may be thrown away. We shall follow the first method, since it is straight forward. Nevertheless, appropriate commutation relations during transition to the quantum theory follow from Dirac brackets. We therefore compute Dirac brackets first and then follow the standard method. The Dirac bracket of two functions $f$ and $g$ in phase space is defined as.

\be \big\{f,g\big\}_{DB} = \big\{f,g\big\}_{PB} - \sum_{ij}\big\{f,\phi_i\big\}_{PB}M^{-1}_{ij}\big\{\phi_j,g\big\}_{PB},\ee
where $M_{ij} = \big\{\phi_i,\phi_j\big\}_{PB}$, possessess an inverse denoted by $M^{-1}_{ij}$. In the present case, the matrix and its inverse are simply
\be M_{ij} = \left(
            \begin{array}{cccc}
              0 & -1 & 0 & 0\\
              1 & 0 & 0 & 0\\
              0 & 0 & 0 & -1\\
              0 & 0 & 1 & 0
            \end{array}
          \right)
~~ \mathrm{and}~~ M^{-1}_{ij} = \left(
            \begin{array}{cccc}
              0 & 1 & 0 & 0\\
              -1 & 0 & 0 & 0\\
              0 & 0 & 0 & 1\\
              0 & 0 & -1 & 0
            \end{array}
          \right)\ee
Therefore, the Dirac bracket reduces to the following form:

\be \big\{f,g\big\}_{DB} = \big\{f,g\big\}_{PB} + \sum_{ij}\epsilon_{ij}\big\{f,\phi_i\big\}_{PB}\big\{\phi_j,g\big\}_{PB},\ee
where $\epsilon_{ij}$ is the Levi-Civita symbol. A straightforward calculation results in:

\be\begin{split} \{z,p_z\}_{DB} &= \{z,p_z\}_{PB} + \epsilon_{11}\{z,\phi_1\}_{PB}\{\phi_1,p_z\}_{PB} + \epsilon_{12}\{z,\phi_1\}_{PB}\{\phi_2,p_z\}_{PB} + \epsilon_{13}\{z,\phi_1\}_{PB}\{\phi_3,p_z\}_{PB}\\
&\hspace{2cm}+ \epsilon_{14}\{z,\phi_1\}_{PB}\{\phi_4,p_z\}_{PB} + \epsilon_{21}\{z,\phi_2\}_{PB}\{\phi_1,p_z\}_{PB} + \epsilon_{22}\{z,\phi_2\}_{PB}\{\phi_2,p_z\}_{PB}\\
&\hspace{2cm} + \epsilon_{23}\{z,\phi_2\}_{PB}\{\phi_3,p_z\}_{PB} + \epsilon_{24}\{z,\phi_2\}_{PB}\{\phi_4,p_z\}_{PB} + \epsilon_{31}\{z,\phi_3\}_{PB}\{\phi_1,p_z\}_{PB}\\
&\hspace{2cm}+ \epsilon_{32}\{z,\phi_3\}_{PB}\{\phi_2,p_z\}_{PB} + \epsilon_{33}\{z,\phi_3\}_{PB}\{\phi_3,p_z\}_{PB} + \epsilon_{34}\{z,\phi_3\}_{PB}\{\phi_4,p_z\}_{PB}\\
&\hspace{2cm} + \epsilon_{41}\{z,\phi_4\}_{PB}\{\phi_1,p_z\}_{PB} + \epsilon_{42}\{z,\phi_4\}_{PB}\{\phi_2,p_z\}_{PB} + \epsilon_{43}\{z,\phi_4\}_{PB}\{\phi_3,p_z\}_{PB}\\
&\hspace{2cm}+ \epsilon_{44}\{z,\phi_4\}_{PB}\{\phi_4,p_z\}_{PB}\\
&= \{z,p_z\}_{PB} = 1,\end{split}\ee
since, $\{\phi_i, p_z\}_{PB} = 0$. Likewise, $\{x,p_x\}_{DB} = \{x,p_x\}_{PB} = 1$, $\{y,p_y\}_{DB} = \{y,p_y\}_{PB} = 1$, $\{w,p_w\}_{DB} = \{w,p_w\}_{PB} = 1$, $\{z,p_x\}_{DB} = \{z,p_x\}_{PB} = 0$, $\{y,p_w\}_{DB} = \{y,p_w\}_{PB} = 0$, $\{p_z,p_x\}_{DB} = \{p_z,p_x\}_{PB} = 0$ and $\{p_y,p_w\}_{DB} = \{p_y,p_w\}_{PB} = 0$. Therefore, the correct implementation of canonical quantization dictates the standard commutation relations, $[\hat{z}, \hat{p}_z] = [\hat{x}, \hat{p}_x] = [\hat{y}, \hat{p}_y] = [\hat{w}, \hat{p}_w] = i\hbar$, $[\hat{z}, \hat{p}_x] = [\hat{y}, \hat{p}_w] = [\hat{p}_z, \hat{p}_x] = [\hat{p}_y, \hat{p}_w] = 0$. The reason behind equality of Dirac bracket and Poisson bracket lies in the fact that, $\phi_2$ and $\phi_4$ strongly vanish. So, while following this prescription, one can throw them away, but not if one follows the first prescription, since in that case it will not be possible to compute $u$'s. Let us now find the phase-space structure of the Hamiltonian following the standard formulation, substituting the four constraints, so that the modified primary Hamiltonian takes the following form,

\be H_{p1} = H_c + u_1(Np_z - \lambda) + u_2p_\lambda + u_3(Np_y - \tau) + u_4p_\tau\ee
In the above, $u_1$, $u_2$, $u_3$ and $u_4$ are Lagrange multipliers, and the Poisson brackets $\{x, p_x\} = \{z, p_z\} = \{\lambda, p_\lambda\} = \{w, p_w\} = \{y, p_y\} = \{\tau, p_\tau\} = 1$, hold \footnote{The Dirac brackets of $H_{p1}$ with any constraint are as follows,

\[ \{\phi_1,H_{p1}\}_{DB} = \{\phi_2,H_{p1}\}_{DB} = \{\phi_3,H_{p1}\}_{DB} = \{\phi_4,H_{p1}\}_{DB} = 0,\]
which is obvious.}.
The requirement that the constraints must remain preserved in time is exhibited in the Poisson brackets $\{\phi_i, H_{p1}\}$ viz,

\be\begin{split} &\dot\phi_1 = \{\phi_1,H_{p1}\} = -N \frac{\partial H_{p1}}{\partial z} - u_2 + \Sigma_{i=1}^2\phi_i\{\phi_1,u_i\},\\
&\dot\phi_2 = \{\phi_2,H_{p1}\} = x - u_1 + \Sigma_{i=1}^2\phi_i\{\phi_2,u_i\},\\
&\dot\phi_3 = \{\phi_3,H_{p1}\} = -N \frac{\partial H_{p1}}{\partial y} - u_4 + \Sigma_{i=1}^2\phi_i\{\phi_3,u_i\},\\
&\dot\phi_4 = \{\phi_4,H_{p1}\} = w - u_3 + \Sigma_{i=1}^2\phi_i\{\phi_4,u_i\}.\end{split}\ee
Now, since constraints should also vanish weakly in the sense of Dirac, so, $\{\phi_1, H_{p1}\} = \dot\phi_1 \approx 0$, sets $u_2 = - N \frac{\partial H_{p1}}{\partial z}$ , $\{\phi_2, H_{p1}\} = \dot\phi_2 \approx 0$, implies $u_1 = x$, $\{\phi_3, H_{p1}\} = \dot\phi_3 \approx 0$, results in $u_4 = - N \frac{\partial H_{p1}}{\partial y}$, and finally, $u_3 = w$ follows from $\{\phi_4, H_{p1}\} = \dot\phi_4 \approx 0$. Imposing these conditions, we express the modified primary Hamiltonian $H_{p2}$ as,

\be\begin{split} H_{p2} &= N\Bigg[ xp_z + wp_y + \frac{\sqrt zp_x^2}{12(3\beta + \gamma)} + \frac{y^2p_w^2}{6\gamma z^{3\over2}} - \frac{3(2\beta + \gamma)w^2zp_x}{4(3\beta + \gamma)y^2} - \frac{3wxp_w}{2z} + \frac{p_\phi^2}{2z^{3\over2}} + U p_\phi + \Bigg\{\frac{U^2}{2}\\
& + \frac{3\gamma(12\beta + 5\gamma)w^4}{16(3\beta + \gamma)y^4} + \frac{3\alpha}{2}\Bigg(\frac{x^2}{z^2} - \frac{w^2}{y^2} \Bigg) + 3\gamma\Bigg(\frac{w^3x}{y^3z} + \frac{3w^2x^2}{4y^2z^2}\Bigg) + V\Bigg\}z^{3\over2} - \frac{\partial H_{p1}}{\partial z}p_\lambda - \frac{\partial H_{p1}}{\partial y}p_\tau\Bigg]\end{split}\ee
We repeat that since constraints should vanish weakly in the sense of Dirac, therefore in view of the Poisson brackets $\{\phi_1, H_{p2}\} = \dot\phi_1 \approx 0$ and $\{\phi_3, H_{p2}\} = \dot\phi_3 \approx 0$, one obtains $p_\lambda = 0$ and $p_\tau = 0$. Thus the Hamiltonian is finally expressed as $H_D = N\mathcal{H_D}$, where,

\be\label{HD}\begin{split} \mathcal{H_D} = &xp_z + wp_y + \frac{\sqrt zp_x^2}{12(3\beta + \gamma)} + \frac{y^2p_w^2}{6\gamma z^{3\over2}} - \frac{3(2\beta + \gamma)w^2zp_x}{4(3\beta + \gamma)y^2} - \frac{3wxp_w}{2z} + \frac{p_\phi^2}{2z^{3\over2}} + U p_\phi\\& + \Bigg\{\frac{U^2}{2}
+ \frac{3\gamma(12\beta + 5\gamma)w^4}{16(3\beta + \gamma)y^4} + \frac{3\alpha}{2}\Bigg(\frac{x^2}{z^2} - \frac{w^2}{y^2} \Bigg) + 3\gamma\Bigg(\frac{w^3x}{y^3z} + \frac{3w^2x^2}{4y^2z^2}\Bigg) + V\Bigg\}z^{3\over2},\end{split}\ee
and in the process diffeomorphic invariance is clearly established. The Hamiltonian so found is identical to the one \eqref{HM2}, found following MHF, i.e., $\mathcal{H_D} = \mathcal{H_M} = \mathcal{H}$ (see appendix A), and thus, equivalence between the two formalisms has been established also in the background of anisotropic Bianchi - 1 model. Now, since $\dot z = Nx$ and $\dot y = Nw$, we have $\dot z p_z + \dot y p_y = N(xp_z + wp_y)$. So, using the expressions of $p_x,~ p_w,~ p_\phi$ in view of equation (\ref{momenta}) and $\mathcal{H}$ from equation (\ref{HD}), it is possible to express the action (\ref{A2}) in the following form:

\be A = \int(\dot zp_z + \dot xp_x + \dot yp_y + \dot wp_w + \dot\phi p_\phi - N\mathcal{H})dt,\ee
which amounts in writing,

\be \label{ADM} A = \int(\dot h_{ij}p^{ij} + \dot K_{ij}\pi^{ij} + \dot\phi p_\phi - N\mathcal{H})dt.\ee
Hence, it has finally been possible to express the much complicated higher-order action \eqref{A} in the anisotropic mini-superspace model \eqref{anisonew}, in the standard canonical (ADM) form. Since we require expressions for all the momenta for semiclassical approximation, therefore, at this end, let us compute $p_z$ and $p_y$ in view of Hamilton's equation (note that $p_x$, $p_w$ and $p_\phi$ are already presented in \eqref{momenta}). Thus we don't need all the Hamilton's equations. In view of the Hamiltonian \eqref{HD}, we therefore, only find the following Hamilton's equations:

\begin{eqnarray}\label{pxdot}\begin{split} \dot {p_x} = - \frac{\partial \mathcal{H_D}}{\partial x} = - p_z + wp_y + \frac{3wp_w}{2z} - U_{,x} p_\phi - &\Bigg\{U_{,x}U + 3\alpha\frac{x}{z^2} + \frac{3\gamma}{z}\Bigg(\frac{w^3}{y^3} + \frac{3w^2x}{2y^2z}\Bigg)\Bigg\}z^{3\over2}\end{split}\\
\label{pwdot}\begin{split} \dot {p_w} &= - \frac{\partial \mathcal{H_D}}{\partial w} = - p_y + \frac{3(2\beta + \gamma)wzp_x}{2(3\beta + \gamma)y^2} + \frac{3xp_w}{2z} - U_{,w} p_\phi - \Bigg\{U_{,w}U + \frac{27(2\beta + \gamma)^2w^3}{4(3\beta + \gamma)y^4}\\
& \hspace{2 in} - 3\alpha\frac{w}{y^2} - 9\beta\frac{w^3}{y^4} + \frac{3\gamma w}{y^2}\Bigg(3\frac{wx}{yz} + 3\frac{x^2}{2z^2} - 3\frac{w^2}{y^2} \Bigg)\Bigg\}z^{3\over2}\end{split}
\end{eqnarray}
Now under the choice $N=1$ and using the expressions of $p_x$, $p_w$ and $p_\phi$ \eqref{momenta}, we can find explicit forms of $p_z$ and $p_y$ from equations (\ref{pxdot}) and \eqref{pwdot} as,

\be\label{pz}\begin{split}& p_z = - \Bigg[3\alpha\frac{x}{z} + 9\beta\Bigg(2\frac{\ddot x}{z} - \frac{\dot x\dot z}{z^2} + \frac{w^2\dot z}{2y^2z} - 2\frac{w^2\dot y}{y^3} + 2\frac{w\dot w}{y^2} \Bigg) + 3\alpha'\dot\phi + 9\beta'\dot\phi\Bigg(2\frac{\dot x}{z} + \frac{w^2}{y^2}\Bigg)\\
& \hspace{0.3 in} + 3\gamma\Bigg(2\frac{\ddot x}{z} - \frac{\dot x\dot z}{z^2} + \frac{3w^2\dot z}{4y^2z} - 3\frac{w^2\dot y}{y^3} + \frac{3w\dot w}{2y^2} - \frac{3w^2x}{4y^2z} + \frac{w^3}{y^3}\Bigg) + 3\gamma'\dot\phi\Bigg(2\frac{\dot x}{z} - \frac{x^2}{2z^2} + \frac{3w^2}{2y^2}\Bigg)\Bigg]\sqrt{z}\\&
p_y = \Bigg[3\alpha w + 9\beta w\Bigg(2\frac{\dot x}{z} + \frac{w^2}{y^2}\Bigg) - 3\gamma w\Bigg(\frac{\ddot w}{w} + \frac{3\dot w\dot z}{2wz} + \frac{3x\dot z}{4z^2} - 2\frac{\dot w\dot y}{wy} - \frac{3\dot x}{2z} - 3\frac{x\dot y}{yz} - \frac{3x^2}{4z^2}\\& \hspace{2.5 in}
 + 3\frac{wx}{yz} - \frac{w^2}{y^2}\Bigg) - 3\gamma'\dot\phi w\Bigg( \frac{\dot w}{w} + \frac{3x}{2z} - \frac{w}{y}\Bigg)\Bigg]\frac{z^{3\over2}}{y^2}.\end{split}\ee
Expressions \eqref{momenta} and \eqref{pz} now constitute the whole set of momenta. Using all these forms of momenta and by putting $x = \dot z$, $w = \dot y$, $z = A^2$, $y = B^2$, it is possible to retrieve the $(^0_0)$ equation of Einstein as presented in (\ref{00}). Also, from the Lagrangian (\ref{L}), we can find the $\phi$ variation equations as presented in equation (\ref{phi}).

\section{Canonical quantization:}

Since the phase-space structure of the action \eqref{A} has been found, canonical quantization of the Hamiltonian \eqref{HD} may now be performed, only after taking care of operator orderings between $\{\hat x, \hat p_x\}$, $\{\hat w, \hat p_w\}$, $\{\hat \phi, \hat p_\phi\}$ etc. suitably. Further, since $U = U(x,y,z,w,\phi)$ contains $\alpha(\phi)$, $\beta(\phi)$ and $\gamma(\phi)$, and there exists coupling between $U$ and $p_\phi$, some operator ordering ambiguities still remain, which may be resolved only after fixing the forms of $\alpha(\phi)$, $\beta(\phi)$ and $\gamma(\phi)$. Note that, both sets of classical de-Sitter solutions \eqref{Sol1} and \eqref{Sol2}, admit identical form of the coupling parameters as well as the potential, viz., we have $V = V_0\phi^2, ~\alpha = \alpha_0\phi^2, ~\beta = \beta_0\phi^2, ~\mathrm{and} ~\gamma = \gamma_0\phi^2$. With this knowledge of the forms of $\alpha(\phi)$, $\beta(\phi)$ and $\gamma(\phi)$, we can perform Weyl symmetric operator ordering between $\hat U$ and $\hat p_\phi$ also. The modified Wheeler-DeWitt equation therefore reads as:

\be\label{quant1}\begin{split} \frac{i\hbar}{\sqrt z}\frac{\partial\Psi}{\partial z} =& - \frac{i\hbar w}{x\sqrt z}\frac{\partial\Psi}{\partial y} - \frac{\hbar^2}{12(3\beta + \gamma)x}\Bigg(\frac{\partial^2}{\partial x^2} + \frac{n}{x}\frac{\partial}{\partial x}\Bigg)\Psi - \frac{\hbar^2y^2}{6\gamma xz^{2}}\Bigg(\frac{\partial^2}{\partial w^2} + \frac{m}{w}\frac{\partial}{\partial w}\Bigg)\Psi\\
& + i\hbar\frac{3(2\beta + \gamma)w^2\sqrt z}{8(3\beta + \gamma)y^2}\Bigg(\frac{2}{x}\frac{\partial\Psi}{\partial x} - \frac{1}{x^2}\Psi\Bigg) + i\hbar\frac{3}{4z^{3\over2}}\Bigg(2w\frac{\partial\Psi}{\partial w} + \Psi\Bigg) - \frac{\hbar^2}{2xz^{2}}\frac{\partial^2\Psi}{\partial\phi^2}\\
& - i\hbar\Bigg[3\frac{\alpha_0}{z^{3\over2}} - \frac{\gamma_0}{x\sqrt z}\Bigg(\frac{x^3}{2z^3} + \frac{w^3}{y^3}\Bigg)\Bigg]\Bigg(2\phi\frac{\partial\Psi}{\partial \phi} + \Psi\Bigg) + \frac{z}{x}\Bigg\{\frac{U^2}{2} + \frac{3\gamma(12\beta + 5\gamma)w^4}{16(3\beta + \gamma)y^4}\\
& + \frac{3\alpha}{2}\Bigg(\frac{x^2}{z^2} - \frac{w^2}{y^2}\Bigg) + 3\gamma\Bigg(\frac{w^3x}{y^3z} + \frac{3w^2x^2}{4y^2z^2}\Bigg) + V\Bigg\}\Psi,\end{split}\ee
where we have introduced operator ordering indices $n$ and $m$ to remove some, but not all the ambiguities. In the above, we have also performed Weyl symmetric operator ordering systematically, as mentioned. Now, to render Schr{\"o}dinger equation-like appearance of the above modified Wheeler-deWitt equation, we undergo a further change of variable, viz. $\sigma = z^{3\over2}$ to obtain,

\be\label{quant2}\begin{split} i\hbar\frac{\partial\Psi}{\partial\sigma} =& - \frac{2i\hbar w}{3x\sigma^{1\over3}}\left(\frac{\partial\Psi}{\partial y}\right) - \frac{\hbar^2}{18(3\beta + \gamma)x}\Bigg(\frac{\partial^2}{\partial x^2} + \frac{n}{x}\frac{\partial}{\partial x}\Bigg)\Psi - \frac{\hbar^2y^2}{9\gamma x\sigma^{4\over3}}\Bigg(\frac{\partial^2}{\partial w^2} + \frac{m}{w}\frac{\partial}{\partial w}\Bigg)\Psi\\
& + i\hbar\frac{(2\beta + \gamma)w^2\sigma^{1\over3}}{4(3\beta + \gamma)y^2}\Bigg(\frac{2}{x}\frac{\partial\Psi}{\partial x} - \frac{1}{x^2}\Psi\Bigg) + \frac{i\hbar}{2\sigma}\Bigg(2w\frac{\partial\Psi}{\partial w} + \Psi\Bigg) - \frac{\hbar^2}{3x\sigma^{4\over3}}\frac{\partial^2\Psi}{\partial\phi^2}\\
& - \frac{2i\hbar}{3}\Bigg[3\frac{\alpha_0}{\sigma} - \frac{\gamma_0}{x\sigma^{1\over3}}\Bigg(\frac{x^3}{2\sigma^2} + \frac{w^3}{y^3}\Bigg)\Bigg]\Bigg(2\phi\frac{\partial\Psi}{\partial \phi} + \Psi\Bigg) + V_e\Psi = \hat H_e\Psi.\end{split}\ee
Clearly, Schr{\"o}dinger equation-like appearance of the modified Wheeler-deWitt equation \eqref{quant2} transpires, due to the fact that the proper volume $\sigma = z^{3\over2} = a^3$,  acts as `internal time parameter'. In the above, $\hat H_e$ is the effective Hamiltonian operator and $V_e = \Big\{\frac{U^2}{3} + \frac{\gamma(12\beta + 5\gamma)w^4}{8(3\beta + \gamma)y^4} + \alpha\Big(\frac{x^2}{\sigma^{4\over3}} - \frac{w^2}{y^2}\Big) + \gamma\Big(2\frac{w^3x}{y^3z} + \frac{3w^2x^2}{2y^2z^2}\Big) + {2\over3}V\Big\}\frac{\sigma^{2\over3}}{x}$ is the effective potential.

\subsection{Hermiticity and probability interpretation:}

To explore hermiticity of the resulting effective Hamiltonian operator, we split $\hat H_e$ appearing in expression \eqref{quant2} as:

\be\label{Hsplit}\begin{split}
&\hat H_e = \hat H_1 + \hat H_2 + \hat H_3 + \hat H_4 + \hat H_5 + \hat H_6 + \hat H_7 + \hat V_e, ~~~~\mathrm{where}\\
&\hat H_1 = - \frac{2i\hbar w}{3x\sigma^{1\over3}}\left(\frac{\partial}{\partial y}\right);\hspace{0.5cm}\hat H_2 = - \frac{\hbar^2}{18(3\beta + \gamma)x}\Bigg(\frac{\partial^2}{\partial x^2} + \frac{n}{x}\frac{\partial}{\partial x}\Bigg);\hspace{0.4cm}\hat H_3 = - \frac{\hbar^2y^2}{9\gamma x\sigma^{4\over3}}\Bigg(\frac{\partial^2}{\partial w^2} + \frac{m}{w}\frac{\partial}{\partial w}\Bigg)\\
&\hat H_4 = i\hbar\frac{(2\beta + \gamma)w^2\sigma^{1\over3}}{4(3\beta + \gamma)y^2}\Bigg(\frac{2}{x}\frac{\partial}{\partial x} - \frac{1}{x^2}\Bigg);\hspace{0.4cm}\hat H_5 = \frac{i\hbar}{2\sigma}\Bigg(2w\frac{\partial}{\partial w} + 1\Bigg);\hspace{0.5cm}\hat H_6 = - \frac{\hbar^2}{3x\sigma^{4\over3}}\left(\frac{\partial^2}{\partial\phi^2}\right)\\
&\hat H_7 = - \frac{2i\hbar}{3}\Bigg[3\frac{\alpha_0}{\sigma} - \frac{\gamma_0}{x\sigma^{1\over3}}\Bigg(\frac{x^3}{2\sigma^2} + \frac{w^3}{y^3}\Bigg)\Bigg]\Bigg(2\phi\frac{\partial}{\partial \phi} + 1\Bigg);\hspace{0.5cm}\hat V_e = V_e
\end{split}\ee
Out of all these, most important are $\hat H_2$ and $\hat H_3$, for the reason that they contain arbitrary operator ordering parameters $n$ and $m$. So, let us consider the second term, integrate it twice and use fall-of condition, to obtain

\be\begin{split} \int (\hat H_2\Psi)^*\Psi dx &= - \frac{\hbar^2}{18(3\beta + \gamma)}\int\Bigg(\frac{1}{x}\frac{\partial^2\Psi^*}{\partial x^2} + \frac{n}{x^2}\frac{\partial\Psi^*}{\partial x}\Bigg)\Psi dx \\&
= - \frac{\hbar^2}{18(3\beta + \gamma)}\int\Bigg(\frac{\Psi}{x}\frac{\partial^2\Psi^*}{\partial x^2} + \frac{n\Psi}{x^2}\frac{\partial\Psi^*}{\partial x}\Bigg) dx\\&
= - \frac{\hbar^2}{18(3\beta + \gamma)}\int \Psi^*\Bigg[\frac{1}{x}\frac{\partial^2\Psi}{\partial x^2} -\left({2+n\over x^2}\right)\frac{\partial\Psi}{\partial x} + \frac{2(n+1)}{x^3}\Psi\Bigg] dx.
\end{split}\ee
Now, only under the choice $n=-1$, one obtains,

\be\begin{split} \int (\hat H_2\Psi)^*\Psi dx = - \frac{\hbar^2}{18(3\beta + \gamma)}\int\Psi^*\Bigg(\frac{1}{x}\frac{\partial^2\Psi}{\partial x^2} - \frac{1}{x^2}\frac{\partial\Psi}{\partial x}\Bigg) dx = \int \Psi^*\hat H_2\Psi dx\end{split}\ee
Thus, $\hat H_2$ is hermitian only if $n=-1$. Likewise, $\hat H_3$ is also hermitian, provided, $m=-1$. In the same manner, rest of the terms are also hermitian. Nevertheless, for completeness, we compute these in appendix B. The effective Hamiltonian operator $\hat H_e$ therefore is hermitian. It is important to mention that the operator ordering parameters $m$ and $n$ have been fixed from physical consideration that the effective Hamiltonian has to be hermitian. The hermiticity of the effective Hamiltonian also allows one to write the continuity equation for $n = m = -1$, as

\be \frac{\partial\rho}{\partial\sigma} + \nabla.\textbf{J} = 0,\ee
where $\rho = \Psi^*\Psi$ and $\textbf{J}=(J_y, J_x, J_w, J_\phi)$ are the probability density and the current density respectively, with $J_y = \frac{4 w}{3x\sigma^{1\over3}}\Psi^*\Psi$, $J_x = \frac{i\hbar}{18(3\beta + \gamma)x}\big(\Psi\Psi^*_{,x} - \Psi^*\Psi_{,x}\big) - \frac{(2\beta + \gamma)w^2\sigma^{1\over3}}{2(3\beta + \gamma)y^2}\frac{\Psi^*\Psi}{x}$, $J_w = \frac{i\hbar y^2}{9\gamma x\sigma^{4\over3}}\big(\Psi\Psi^*_{,w} - \Psi^*\Psi_{,w}\big) - \frac{w}{\sigma}\Psi^*\Psi$, $J_\phi = \frac{i\hbar}{3x\sigma^{4\over3}}\big(\Psi\Psi^*_{,\phi} - \Psi^*\Psi_{,\phi}\big) + \frac{4}{3}\big[3\frac{\alpha_0}{\sigma} - \frac{\gamma_0}{x\sigma^{1\over3}}\big(\frac{x^3}{2\sigma^2} + \frac{w^3}{y^3}\big)\big]\phi\Psi^*\Psi$.
Once again we mention that the continuity equation in the above standard form is found only fixing the factor ordering indices to $n = m = -1$. In the process, the arbitrariness of the factor ordering indices has been removed from physical argument.\\

\subsection{Semiclassical solution under WKB approximation:}

Semiclassical approximation is a method of finding an approximate wavefunction associated with a quantum equation. While unitarity proves the viability of a quantum equation in quantum domain, an approximate classically allowed wavefunction obtained following an appropriate semiclassical approximation, authenticates it in the classical regime. If the integrand in the exponent of the semiclassical wavefunction is imaginary, then the behaviour of the approximate wave function is oscillatory, and falls within the classically allowed region. Otherwise it is classically forbidden. In this connection, let us recall Hartle criterion for the selection of classical trajectories \cite{Hartle}. If the wave function of the universe is strongly peaked, then it admits correlations among the geometrical and matter degrees of freedom, and therefore the emergence of classical trajectories of the universe is expected. Otherwise, such correlations are lost. In this subsection, we therefore check if the above quantum equation admits a feasible semiclassical wavefunction. Instead of considering the time dependent Schr{\"o}dinger equation (\ref{quant2}), let us, for the sake of simplicity and convenience, consider the time independent form of the equation (\ref{quant1}) to compute semiclassical wavefunction in the standard WKB method. We therefore express it as,

\be\label{semi}\begin{split}  & - \frac{\hbar^2\sqrt z}{12(3\beta + \gamma)}\Bigg(\frac{\partial^2}{\partial x^2} + \frac{n}{x}\frac{\partial}{\partial x}\Bigg)\Psi - \frac{\hbar^2y^2}{6\gamma z^{3\over2}}\Bigg(\frac{\partial^2}{\partial w^2} + \frac{m}{w}\frac{\partial}{\partial w}\Bigg)\Psi  - \frac{\hbar^2}{2z^{3\over2}}\frac{\partial^2\Psi}{\partial\phi^2} - i\hbar x\frac{\partial\Psi}{\partial z} - i\hbar w\frac{\partial\Psi}{\partial y} \\
&~~~~~~ + i\hbar\frac{3(2\beta + \gamma)w^2 z}{4(3\beta + \gamma)y^2}\Big(\frac{\partial\Psi}{\partial x}\Big) + i\hbar\frac{3wx}{2z}\Big(\frac{\partial\Psi}{\partial w}\Big) - 2i\hbar\Bigg[3\alpha_0\frac{x}{z} - \gamma_0\Bigg(\frac{x^3}{2z^3} + \frac{w^3}{y^3}\Bigg)\Bigg]\phi\frac{\partial\Psi}{\partial \phi} + \mathcal {V}\Psi = 0,\end{split}\ee
where $\mathcal {V} = \Big[\frac{U^2}{2} + \frac{3\gamma(12\beta + 5\gamma)w^4}{16(3\beta + \gamma)y^4} + \frac{3\alpha}{2}\Big(\frac{x^2}{z^2} - \frac{w^2}{y^2}\Big) + \gamma\Big(3\frac{w^3x}{y^3z} + \frac{9w^2x^2}{4y^2z^2}\Big) + V\Big]z^{3\over2} - i\hbar\frac{3(2\beta + \gamma)w^2 z}{8(3\beta + \gamma)y^2x} + i\hbar\frac{3x}{4z} - i\hbar\Big[3\alpha_0\frac{x}{z} - \gamma_0\Big(\frac{x^3}{2z^3} + \frac{w^3}{y^3}\Big)\Big].$ The above equation is essentially a time independent Schr{\"o}dinger equation consisting of five variables $x, z, w, y$ and $\phi$ and therefore, customarily, let us seek the solution of equation (\ref{semi}) as,
\be\label{psi} \psi = \psi_0(x,z,w,y,\phi)e^{\frac{i}{\hbar}S(x,z,w,y,\phi)},\ee
where, the amplitude $\psi_0$ varies slowly with respect to the phase $S$. As usual, we expand $S$ in power series of $\hbar$, in the following manner:

\be\label{S} S = S_0(x,z,w,y,\phi) + \hbar S_1(x,z,w,y,\phi) + \hbar^2S_2(x,z,w,y,\phi) + .... \ .\ee
As a result one can compute,

\be\label{derivative}\begin{split} &\Psi_{,x} = \psi_{0,x}e^{\frac{i}{\hbar}S} + \frac{i}{\hbar}\Big[S_{0,x} + \hbar S_{1,x} + \hbar^2S_{2,x} + \mathcal{O}(\hbar)\Big]\psi_0e^{\frac{i}{\hbar}S};\\
&\Psi_{,xx} = \Psi_{0,xx}e^{\frac{i}{\hbar}S} + 2\frac{i}{\hbar}\Big[S_{0,x} + \hbar S_{1,x} + \hbar^2S_{2,x} + \mathcal{O}(\hbar)\Big]\psi_{0,x}e^{\frac{i}{\hbar}S} + \frac{i}{\hbar}\Big[S_{0,xx} + \hbar S_{1,xx} + \hbar^2S_{2,xx} + \mathcal{O}(\hbar)\Big]\psi_0e^{\frac{i}{\hbar}S}\\
&\hspace{1.2cm} - \frac{1}{\hbar^2}\Big[S_{0,x}^2 + \hbar^2 S_{1,x}^2 + \hbar^4S_{2,x}^4 + 2\hbar S_{0,x}S_{1,x} + 2\hbar^2 S_{0,x}S_{2,x} + 2\hbar^3 S_{1,x}S_{2,x} + \mathcal{O}(\hbar)\Big]\psi_0e^{\frac{i}{\hbar}S};\\
&\Psi_{,w} = \psi_{0,w}e^{\frac{i}{\hbar}S} + \frac{i}{\hbar}\Big[S_{0,w} + \hbar S_{1,w} + \hbar^2S_{2,w} + \mathcal{O}(\hbar)\Big]\psi_0e^{\frac{i}{\hbar}S};\\
&\Psi_{,ww} = \Psi_{0,ww}e^{\frac{i}{\hbar}S} + 2\frac{i}{\hbar}\Big[S_{0,w} + \hbar S_{1,w} + \hbar^2S_{2,w} + \mathcal{O}(\hbar)\Big]\psi_{0,w}e^{\frac{i}{\hbar}S} + \frac{i}{\hbar}\Big[S_{0,ww} + \hbar S_{1,ww} + \hbar^2S_{2,ww} + \mathcal{O}(\hbar)\Big]\psi_0e^{\frac{i}{\hbar}S}\\
&\hspace{1.2cm} - \frac{1}{\hbar^2}\Big[S_{0,w}^2 + \hbar^2 S_{1,w}^2 + \hbar^4S_{2,w}^4 + 2\hbar S_{0,w}S_{1,w} + 2\hbar^2 S_{0,w}S_{2,w} + 2\hbar^3 S_{1,w}S_{2,w} + \mathcal{O}(\hbar)\Big]\psi_0e^{\frac{i}{\hbar}S};\\
&\Psi_{,\phi} = \psi_{0,\phi}e^{\frac{i}{\hbar}S} + \frac{i}{\hbar}\Big[S_{0,\phi} + \hbar S_{1,\phi} + \hbar^2S_{2,\phi} + \mathcal{O}(\hbar)\Big]\psi_0e^{\frac{i}{\hbar}S};\\
&\Psi_{,\phi\phi} = \Psi_{0,\phi\phi}e^{\frac{i}{\hbar}S} + 2\frac{i}{\hbar}\Big[S_{0,\phi} + \hbar S_{1,\phi} + \hbar^2S_{2,\phi} + \mathcal{O}(\hbar)\Big]\psi_{0,\phi}e^{\frac{i}{\hbar}S} + \frac{i}{\hbar}\Big[S_{0,\phi\phi} + \hbar S_{1,\phi\phi} + \hbar^2S_{2,\phi\phi} + \mathcal{O}(\hbar)\Big]\psi_{0}e^{\frac{i}{\hbar}S}\\
&\hspace{1.2cm} - \frac{1}{\hbar^2}\Big[S_{0,\phi}^2 + \hbar^2 S_{1,\phi}^2 + \hbar^4S_{2,\phi}^4 + 2\hbar S_{0,\phi}S_{1,\phi} + 2\hbar^2 S_{0,\phi}S_{2,\phi} + 2\hbar^3 S_{1,\phi}S_{2,\phi} + \mathcal{O}(\hbar)\Big]\psi_{0}e^{\frac{i}{\hbar}S};\\
&\Psi_{,z} = \psi_{0,z}e^{\frac{i}{\hbar}S} + \frac{i}{\hbar}\Big[S_{0,z} + \hbar S_{1,z} + \hbar^2S_{2,z} + \mathcal{O}(\hbar)\Big]\psi_{0}e^{\frac{i}{\hbar}S};\\
&\Psi_{,y} = \psi_{0,y}e^{\frac{i}{\hbar}S} + \frac{i}{\hbar}\Big[S_{0,y} + \hbar S_{1,y} + \hbar^2S_{2,y} + \mathcal{O}(\hbar)\Big]\psi_{0}e^{\frac{i}{\hbar}S}.\end{split}\ee
\noindent
In the above, `comma' everywhere in the suffix represents derivative. Now plugging in all these expressions (\ref{derivative}) in equation (\ref{semi}), and equating the coefficients of different powers of $\hbar$ to zero, we obtain the set of equations (upto second order), as follows:

\be\label{semi1}\begin{split}\frac{\sqrt z}{12(3\beta + \gamma)}S_{0,x}^2 + \frac{y^2}{6\gamma z^{3\over2}}S_{0,w}^2 + \frac{1}{2z^{3\over2}}S_{0,\phi}^2 +& xS_{0,z} + wS_{0,y} - \frac{3(2\beta + \gamma)w^2 z}{4(3\beta + \gamma)y^2}S_{0,x} - \frac{3wx}{2z}S_{0,w}\\
& + 2\Bigg[3\alpha_0\frac{x}{z} - \gamma_0\Bigg(\frac{x^3}{2z^3} + \frac{w^3}{y^3}\Bigg)\Bigg]\phi S_{0,\phi} + V_1= 0.\end{split}\ee
\be\label{semi2}\begin{split} -&\frac{\sqrt z}{12(3\beta + \gamma)}\Big[\Big(i S_{0,xx} - 2S_{0,x}S_{1,x} + \frac{i}{x}n S_{0,x}\Big)\psi_0 + 2iS_{0,x}\psi_{0,x} \Big] - \frac{y^2}{6\gamma z^{3\over2}}\Big[\Big(i S_{0,ww} - 2S_{0,w}S_{1,w}\\
&~~~~ + \frac{i}{w}m S_{0,w}\Big)\psi_0 + 2iS_{0,w}\psi_{0,w}\Big] - \frac{1}{2z^{3\over2}}\Big[\Big(i S_{0,\phi\phi} - 2S_{0,\phi}S_{1,\phi}\Big)\psi_0 + 2iS_{0,\phi}\psi_{0,\phi}\Big] + x S_{1,z}\psi_0\\
&~~~~ - ix\psi_{0,z} + wS_{1,y}\psi_0 - iw\psi_{0,y} - \frac{3(2\beta + \gamma)w^2 z}{4(3\beta + \gamma)y^2}\Big[\Big(S_{1,x} + \frac{i}{2x}\Big)\psi_0 - i\psi_{0,x}\Big] - \frac{3wx}{2z}\Big[\Big(S_{1,w}\\
&~~~~ - \frac{i}{2w}\Big)\psi_0 - i\psi_{0,w}\Big] + \Bigg[3\alpha_0\frac{x}{z} - \gamma_0\Bigg(\frac{x^3}{2z^3} + \frac{w^3}{y^3}\Bigg)\Bigg]\Big[\Big(2\phi S_{1,\phi} - i\Big)\psi_0 - 2i\phi\psi_{0,\phi}\Big] = 0.\end{split}\ee
\be\label{semi3}\begin{split} -&\frac{\sqrt z}{12(3\beta + \gamma)}\Big[\Big(i S_{1,xx} - S_{1,x}^2 - 2S_{0,x}S_{2,x} + \frac{i}{x}n S_{1,x}\Big)\psi_0 + \psi_{0,xx} + 2iS_{1,x}\psi_{0,x} + \frac{n}{x}\psi_{0,x}\Big]\\
&~~ - \frac{y^2}{6\gamma z^{3\over2}}\Big[\Big(i S_{1,ww} - S_{1,w}^2 - 2S_{0,w}S_{2,w} + \frac{i}{w}m S_{1,w}\Big)\psi_0 + \psi_{0,ww} + 2iS_{1,w}\psi_{0,w} + \frac{m}{w}\psi_{0,w}\Big]\\
&~~ - \frac{1}{2z^{3\over2}}\Big[\Big(i S_{1,\phi\phi} - S_{1,\phi}^2 - 2S_{0,\phi}S_{2,\phi}\Big)\psi_0 + \psi_{0,\phi\phi} + 2iS_{1\phi}\psi_{0,\phi}\Big] - \frac{3(2\beta + \gamma)w^2 z}{4(3\beta + \gamma)y^2}S_{2,x}\\
&~~ + x S_{2,z} + wS_{2,y} - \frac{3wx}{2z}S_{2,w} + 2\Bigg[3\alpha_0\frac{x}{z} - \gamma_0\Bigg(\frac{x^3}{2z^3} + \frac{w^3}{y^3}\Bigg)\Bigg]\phi S_{2,\phi} = 0,\end{split}\ee
where, $V_1 = \Big[\frac{U^2}{2} + \frac{3\gamma(12\beta + 5\gamma)w^4}{16(3\beta + \gamma)y^4} + \frac{3\alpha}{2}\Big(\frac{x^2}{z^2} - \frac{w^2}{y^2}\Big) + \gamma\Big(3\frac{w^3x}{y^3z} + \frac{9w^2x^2}{4y^2z^2}\Big) + V\Big]z^{3\over2}.$ These equations \eqref{semi1}-\eqref{semi3} are to be solved successively to find $S_0,\; S_1$ and $S_2$ and so on. Now identifying $S_{0,x}$ as $p_x$, $S_{0,w}$ as $p_w$, $S_{0,\phi}$ as $p_\phi$, $S_{o,z}$ as $p_z$ and $S_{0,y}$ as $p_y$; the classical Hamiltonian constraint equation ${\mathcal{H}} = 0$, presented in equation (\ref{HD}), may be recovered from equation (\ref{semi1}), which is therefore identified as the Hamilton-Jacobi equation. Hence the Hamiltonian (\ref{HD}) successfully passes the first consistency check. The Hamilton-Jacobi function, $S_0(x,z,w,y,\phi)$ therefore is expressed as,

\be\label{S0} S_0 = \int p_x dx + \int p_z dz + \int p_w dw + \int p_y dy + \int p_\phi d\phi, \ee
apart from a constant of integration which may be absorbed in $\psi_0$. It is possible to evaluate the integrals in the above expression, using the classical solution \eqref{Sol1} together with the definitions of momenta presented in \eqref{momenta} and \eqref{pz}, keeping the relations $x = \frac{\dot z}{N}$ and $w = \frac{\dot y}{N}$, where, $z = a^2 = \big(AB^2\big)^{2\over3}$ and $y = \big({B\over A}\big)^{1\over3}$, in mind. For the present purpose, we fix the gauge $N = 1$. Thus, the expressions of momenta are found as,

\be\begin{split}
&p_x = 6\lambda^2\big(54\beta_0 + 19\gamma_0\big)\phi_0^2\big(A_0B_0^2\big)^{2\over3} = \mathrm{const.}\hspace{1cm} p_z = - 6\lambda\big(\alpha_0 + 4\gamma_0\lambda^2\big)\phi_0^2\big(A_0B_0^2\big)^{1\over3} = \mathrm{const.}\\
&p_w = \frac{192\gamma_0\lambda^5\phi_0^2{B_0}^{10\over3}}{{A_0}^{1\over3}}\left({1\over{w^3}}\right)\hspace{1cm} p_y = - \frac{6\lambda\big(\alpha_0 + 108\beta_0\lambda^2 + 56\gamma_0\lambda^2\big)\phi_0^2{B_0}^{10\over3}}{{A_0}^{1\over3}}\left({1\over{y^3}}\right)\hspace{0.45cm}\\
&p_\phi = - \lambda\big(24\alpha_0 - 48\gamma_0\lambda^2 + 1\big)A_0{B_0}^2\left(\frac{{\phi_0}^6}{\phi^5}\right),
\end{split}\ee
and hence the integrals in (\ref{S0}) are evaluated as,

\be\begin{split}
&\int p_xdx = 6\lambda^2\big(54\beta_0 + 19\gamma_0\big)\phi_0^2\big(A_0B_0^2\big)^{1\over3}x = 24\lambda^3\big(54\beta_0 + 19\gamma_0\big)\phi_0^2\big(A_0B_0^2\big)^{1\over3}z;\\
&\int p_zdz = - 6\lambda\big(\alpha_0 + 4\gamma_0\lambda^2\big)\phi_0^2\big(A_0B_0^2\big)^{1\over3}z;
\int p_wdw = - \frac{96\gamma_0\lambda^5\phi_0^2{B_0}^{10\over3}}{{A_0}^{1\over3}w^2} = - 24\gamma_0\lambda^3\phi_0^2\big(A_0B_0^2\big)^{1\over3}z;\\
&\int p_ydy = \frac{3\lambda\big(\alpha_0 + 108\beta_0\lambda^2 + 56\gamma_0\lambda^2\big)\phi_0^2{B_0}^{10\over3}}{{A_0}^{1\over3}y^2} = 3\lambda\big(\alpha_0 + 108\beta_0\lambda^2 + 56\gamma_0\lambda^2\big)\phi_0^2\big(A_0B_0^2\big)^{1\over3}z;\\
&\int p_\phi d\phi = \frac{\lambda\left(6\alpha_0 - 12\gamma_0\lambda^2 + {1\over4}\right)A_0{B_0}^2{\phi_0}^6}{\phi^4} = \lambda\left(6\alpha_0 - 12\gamma_0\lambda^2 + {1\over4}\right)\phi_0^2\big(A_0B_0^2\big)^{1\over3}z.
\end{split}\ee

\noindent
Hence, explicit form of $S_0$ is,

\be S_0 = \lambda\left[3\Big(\alpha_0 + 540\beta_0\lambda^2 + 188\gamma_0\lambda^2\Big) + {1\over4}\right]\phi_0^2\big(A_0B_0^2\big)^{1\over3}z.\ee
In view of the classical solutions (\ref{Sol1}), it is also possible to compute the zeroth order on-shell action (\ref{A2}) as:

\be\begin{split} A_{cl} &= \int{\lambda\over2}\left[3\Big(\alpha_0 + 486\beta_0\lambda^2 + 178\gamma_0\lambda^2\Big) + {1\over4} - \frac{V_0}{2\lambda^2}\right]\phi_0^2\big(A_0B_0^2\big)^{1\over3}dz\\
& = {\lambda\over2}\left[3\Big(\alpha_0 + 486\beta_0\lambda^2 + 178\gamma_0\lambda^2\Big) + {1\over4} - \frac{V_0}{2\lambda^2}\right]\phi_0^2\big(A_0B_0^2\big)^{1\over3}z\end{split}\ee
At first sight, the expressions of $S_0$ and $A_{cl}$ look different, but if we use the expressions of $\alpha_0$ and $\gamma_0$ from classical solution \eqref{Sol1}, then we have

\be S_0 = A_{cl} = \lambda\left[{1\over162} - 72\beta_0\lambda^2 - \frac{38V_0}{81\lambda^2}\right]\phi_0^2\big(A_0B_0^2\big)^{1\over3}z.\ee
Since classical on-shell action is exactly the same as the Hamilton-Jacobi function, so the Hamiltonian (\ref{HD}) also passes through the second consistency check. Here, we mention that earlier in one case \cite{Chandra}, we found that although $S_0 = A_{cl}$, the Hamilton-Jacobi function did not satisfy (supposed-to-be) the Hamilton-Jacobi equation. For this reason, we perform the above computation, and one can check that in the present case, $S_0$ satisfies the Hamilton-Jacobi equation \eqref{semi1} identically. One can now express the semiclassical wave function as,

\be \psi = \psi_{01} e^{\frac{i}{\hbar}\lambda\big[{1\over162} - 72\beta_0\lambda^2 - \frac{38V_0}{81\lambda^2}\big]\phi_0^2(A_0B_0^2)^{1\over3}z}.\ee
It is extremely difficult to solve the first order equation (\ref{semi2}), exactly. However, since everything may be expressed in terms of $z$, a little algebra, neglecting some derivative terms associated with slowly varying amplitude, reveals that one can in principle, express (\ref{semi2}) in the form $S_1 = iG_1(z)$ on the solutions (\ref{Sol1}). Therefore up to the first order approximation, the wavefunction may be expressed as:

\be \label{psi}\psi = \psi_{02} e^{\frac{i}{\hbar}\lambda\big[{1\over162} - 72\beta_0\lambda^2 - \frac{38V_0}{81\lambda^2}\big]\phi_0^2(A_0B_0^2)^{1\over3}z}, \hspace{0.5cm}\mathrm{where},\hspace{0.5cm}\psi_{02} = \psi_{01}e^{-G_1(z)}.\ee
The above form of the semiclassical wavefunction \eqref{psi}, obtained upto first-order approximation, retains the  oscillatory behaviour of the wave function unaltered, while modifies the pre-factor only. This indicates a classically allowed region, and the wavefunction is therefore strongly peaked about a set of classical de-Sitter solutions (\ref{Sol1}). Thus, a clear correspondence between the quantum and the classical domains is established, resulting in a viable quantum theory \cite{Hartle}. So altogether, for the action (\ref{A}), the Hamiltonian (\ref{HD}) is particularly well-behaved.

\section{Inflation under slow roll approximation:}

Although there are alternatives \cite{alt1, alt2, alt3}, inflation is indeed the mainstream choice, since perhaps it is the simplest scenario, which solves the horizon and the flatness problems and also elegantly explains the origin of the seeds of perturbation singlehandedly. Inflation is essentially a quantum theory of perturbation, which occurred at the sub-Planckian epoch, when gravity became classical. So the energy scale of the background must be much below Planck scale, sometime between between $10^{-36} - 10^{-26}$s, although there are few exceptions. It is also important to mention that the semiclassical approximation performed above is also validated if the energy scale of inflation is sub-Planckian. Since the present quantum theory admits a viable semiclassical approximation, therefore most of the important physics may be extracted from the classical action itself.  Inflation therefore may be studied in view of the classical field equations \eqref{00} and \eqref{phi}, which are translated in the transformed matric \eqref{anisonew} to,

\be\label{slow00}\begin{split} \frac{1}{2}{\dot\phi}^2 + V =& 6\alpha\big(\dot\xi^2 - \dot\chi^2\big) + 36\beta\big(2\dot\xi\dddot\xi + 6\dot\xi^2\ddot\xi - 4\dot\chi^2\ddot\xi - \ddot\xi^2 + 4\dot\xi\dot\chi\ddot\chi - 6\dot\xi^2\dot\chi^2 - 3\dot\chi^4\big)\\
& + 6\gamma\big(4\dot\xi\dddot\xi + 2\dot\chi\dddot\chi + 12\dot\xi^2\ddot\xi - 6\dot\chi^2\ddot\xi - 2\ddot\xi^2 + 12\dot\xi\dot\chi\ddot\chi - \ddot\chi^2 - 9\dot\xi^2\dot\chi^2 - 18\dot\chi^4\big)\\
& + 6\alpha'\dot\phi\dot\xi + 72\beta'\dot\phi\dot\xi\big(\ddot\xi + 2\dot\xi^2 + \dot\chi^2\big) + 12\gamma'\dot\phi\big(2\dot\xi\ddot\xi + 3\dot\xi^3 + 6\dot\xi\dot\chi^2 + \dot\chi\ddot\chi\big)\end{split}\ee

\be\label{slowphi}\begin{split} \ddot\phi + 3\dot\xi\dot\phi =& - V' + 6\alpha'\big(\ddot\xi + 2\dot\xi^2 + \dot\chi^2\big) + 36\beta'\big(\ddot\xi^2 + 4\dot\xi^2\ddot\xi + 2\dot\chi^2\ddot\xi + 4\dot\xi^4 + 4\dot\xi^2\dot\chi^2 + \dot\chi^4\big)\\
& + 6\gamma'\big(2\ddot\xi^2 + \ddot\chi^2 + 6\dot\xi^2\ddot\xi + 6\dot\chi^2\ddot\xi + 6\dot\xi\dot\chi\ddot\chi + 6\dot\xi^4 + 15\dot\xi^2\dot\chi^2 + 6\dot\chi^4\big).\end{split}\ee
Making the following replacements: $\dot\xi = H$, $\dot\alpha = \alpha'\dot\phi$, $\dot\beta = \beta'\dot\phi$ and $\dot\gamma = \gamma'\dot\phi$ and using the relation $\dot\chi = \frac{cH}{\sqrt 3}$ in view of the solutions \eqref{Sol2} of set II, the above two equations (\ref{slow00}) and (\ref{slowphi}) are expressed as,

\be\label{slow001}\begin{split} \frac{1}{2}{\dot\phi}^2 + V =& 2\alpha (3 - c^2)H^2 + 12\beta\big\{6H\ddot H + 18H^2\dot H - \dot H^2 - c^2(c^2 + 6)H^4\big\}\\
& + 2\gamma\big\{(c^2 + 6)\big(2H\ddot H + 6H^2\dot H - \dot H^2\big) - 3c^2(2c^2 + 3)H^4\big\} + 6H\dot\alpha\\
& + 24H\dot\beta\big\{3\dot H + (c^2 + 6)H^2\big\}  + 4H\dot\gamma\big\{(c^2 + 6)\dot H + 3(2c^2 + 3)H^2\big\}\end{split}\ee

\be\label{slowphi1}\begin{split} \ddot\phi + 3H\dot\phi =& - V' + 2\alpha'\big\{3\dot H + (c^2 + 6)H^2\big\} + 4\beta'\big\{3\dot H + (c^2 + 6)H^2\big\}^2\\
& + 2\gamma'\big\{(c^2 + 6)\dot H^2 + 6(2c^2 + 3)H^2\dot H + (c^2 + 6)(2c^2 + 3)H^4\big\}\end{split}\ee
Now, we restrict ourselves to $\alpha = \alpha_0$ = const, $\beta = \beta_0\phi^2$ and $\gamma = \gamma_0\phi^2$, so that, $\dot \alpha = 0$ and $\dot\gamma = \frac{\gamma_0}{\beta_0}\dot\beta$, while $\ddot\beta = \beta''\dot\phi^2 + \beta'\ddot\phi$. The presence of additional degree of freedom $\beta(\phi)$, requires to impose two additional conditions, viz. $4|\dot\beta|H \ll 1$ and $|\ddot\beta| \ll |\dot\beta|H$ \cite{Sodaprd}, in addition to the standard slow-roll conditions of minimally coupled single-field inflation, viz. $\dot\phi^2 \ll V$ and $|\ddot\phi| \ll 3H|\dot\phi|$. Instead of standard slow-roll parameters, it is now customary to introduce a combined hierarchy of Hubble and other flow parameters \cite{Schwarzplb, Leach, Schwarzjcap, Sodajcap}, in the following manner. The background evolution at first is described by a set of horizon flow functions (the behaviour of Hubble distance during inflation) starting from,

\be\label{hier1} \epsilon_0 = \frac{d_H}{d_{H_i}}, ~~~~\mathrm{where}~~ d_H = H^{-1}.\ee
In the above, $d_H$ is commonly called the Hubble distance. In what follows, we define the hierarchy of functions systematically.

\be\label{hier2} \epsilon_{l+1} = \frac{d \ln |\epsilon_l|}{d\mathrm{N}},~~ l \ge 0.\ee
One can now compute $\epsilon_1 = \frac{d \ln d_H}{d{\mathrm{N}}}$, following the definition of the number of e-folds $\mathrm{N} = \ln{\big(\frac{a}{a_i}\big)}$, which implies $\dot {\mathrm{N}} = H$. The logarithmic change of Hubble distance per e-fold expansion $\mathrm{N}$ is the first slow-roll parameter $\epsilon_1 = \dot d_H = - \frac{\dot H}{H^2}$. In view of the above hierarchy, we can compute $\epsilon_2 = \frac{d \ln \epsilon_1}{d\mathrm{N}} = \frac{1}{H}\frac{\dot\epsilon_1}{\epsilon_1}$, implying $\epsilon_1\epsilon_2 = d_H\ddot d_H = - \frac{1}{H^2}\Big(\frac{\ddot H}{H} - 2\frac{\dot H^2}{H^2}\Big)$. It is further possible to compute higher slow-roll parameters in the same manner. Treating cosmic time as the evolution parameter, equation (\ref{hier2}) essentially defines a flow in space, which is described by the equation of motion,

\be\label{hier3} \epsilon_0\dot\epsilon_l - \frac{1}{d_{H_i}}\epsilon_l\epsilon_{l-1} = 0,~~ l \ge 0.\ee
Using the definition (\ref{hier1}), the results obtained from the hierarchy defined in (\ref{hier2}) may now be extracted from relation (\ref{hier3}). The additional degree of freedom appearing here, as mentioned, requires yet another hierarchy of flow parameters, which is:

\be \delta_1 = 4\dot\beta H \ll 1,~~ \delta_{i+1} = \frac{d \ln |\delta_i|}{d \ln a},~~ \mathrm{with},~~ i \ge 1.\ee
Note that, for $i = 1, \delta_2 = \frac{d \ln |\delta_1|}{d\mathrm{N}} = \frac{1}{\delta_1}\frac{\dot\delta_1}{\dot {\mathrm{N}}}$, and $\delta_1\delta_2 = \frac{4}{H}(\ddot\beta H + \dot\beta\dot H)$, etc. In analogy to the standard slow-roll approximation. The slow-roll conditions therefore read as, $|\epsilon_i| \ll 1$ and $|\delta_i| \ll 1$, and the field equations (\ref{slow001}) and (\ref{slowphi1}) may therefore be expressed as,

\be\label{slow002}\begin{split} \frac{1}{2}{\dot\phi}^2 + V =& 2\alpha_0 (3 - c^2)H^2 + 12\beta\big\{3\epsilon_1(3\epsilon_1 - 2\epsilon_2 - 6) - c^2(c^2 + 6)\big\}H^4\\
& + 2\gamma\big\{(c^2 + 6)\epsilon_1(3\epsilon_1 - 2\epsilon_2 - 6) - 3c^2(2c^2 + 3)\big\}H^4\\
& + 6\delta_1(c^2 + 6 - 3\epsilon_1)H^3 + \frac{\gamma_0}{\beta_0}\delta_1\big\{3(2c^2 + 3) - (c^2 + 6)\epsilon_1\big\}H^3\end{split}\ee

\be\label{slowphi2}\begin{split} \ddot\phi + 3H\dot\phi =& - V' + 4\beta'(c^2 + 6 -3\epsilon_1)^2H^4 + 2\gamma'\big\{(c^2 + 6)\epsilon_1^2 - 6(2c^2 + 3)\epsilon_1 + (c^2 + 6)(2c^2 + 3)\big\}H^4\end{split}\ee
respectively, which are approximated to:

\be\label{slow003} V \simeq 2\alpha_0 (3 - c^2)H^2 - 12\beta c^2(c^2 + 6)H^4 - 6\gamma c^2(2c^2 + 3)H^4,\ee

\be\label{slowphi3} 3H\dot\phi \simeq - V' + 4\beta'(c^2 + 6)^2H^4 + 2\gamma'(c^2 + 6)(2c^2 + 3)H^4.\ee
Further, choosing, $2\beta_0(c^2 + 6) + \gamma_0 (2c^2 + 3) = 0$, the above equations may be drastically simplified to,

\be\label{slow004} V \simeq 2\alpha_0 (3 - c^2)H^2 ~~~~\mathrm{and}~~~~ 3H\dot\phi \simeq - V',\ee
which take the form of single scalar field model. The above pair of equations are combined to yield,

\be \frac{H}{\dot\phi} = - \frac{3}{2\alpha_0(3 - c^2)}\frac{V}{V'}\ee
The number of e-folds therefore can now be computed using the following relation,

\be \mathrm{N}(\phi) \simeq \int_{t_i}^{t_f} H dt \simeq \int_{\phi_i}^{\phi_f} \frac{H}{\dot\phi} d\phi \simeq \frac{3}{2\alpha_0(3 - c^2)}\int_{\phi_f}^{\phi_i} \frac{V}{V'} d\phi\ee
where, $\phi_i$ and $\phi_f$ denote the value of the scalar fields at the beginning $(t_i)$ and at the end $(t_f)$ of inflation, respectively. Finally, if we choose quadratic potential in the form, $V(\phi) = V_1 + V_0\phi^2$, in accord to the classical solution, the number of e-folds reads as,

\be \mathrm{N}(\phi) = \frac{3}{2\alpha_0(3 - c^2)}\int_{\phi_f}^{\phi_i} \frac{V_1 + V_0\phi^2}{2V_0\phi} d\phi = \frac{3}{4\alpha_0(3 - c^2)}\Bigg[\frac{V_1}{V_0}\ln\frac{\phi_i}{\phi_f} + \frac{1}{2}(\phi_i^2 - \phi_f^2)\Bigg].\ee
It is now possible to compute all other slow-roll parameters, which we do as an example. Setting $\alpha_0 = \frac{M_{P}^2}{2}$, $M_P = (8\pi G)^{-1}$ being the Planck's mass, while the constant, $c =1$, and under the following choice of numerical values: $\frac{V_1}{V_0} = -500M_{P}^2$, $\phi_i = 32.74 M_{P}$, inflation halts ($\epsilon_f = 1$) for $\phi_f = 21.79 M_{P}$, after $\mathrm{N} = 71$ e-folds of expansion. In view of equation (\ref{slow004}), the slow-roll parameters take the following numerical values,

\be \epsilon = - \frac{\dot H}{H^2} = \frac{1}{3} \alpha_0 (3 - c^2) \Bigg(\frac{V'}{V}\Bigg)^2 = \frac{1}{3} \alpha_0 (3 - c^2) \Bigg(\frac{2V_0\phi_i}{V_1 + V_0\phi_i^2}\Bigg)^2 = \frac{4}{3} \alpha_0 (3 - c^2) \frac{\phi_i^2}{\Big(\frac{V_1}{V_0} + \phi_i^2\Big)^2} = 0.00437\ee

\be \eta = \frac{\ddot\phi}{H\dot\phi} = - \frac{2}{3} \alpha_0 (3 - c^2) \frac{V''}{V} = - \frac{2}{3} \alpha_0 (3 - c^2) \frac{2V_0}{V_1 + V_0\phi_i^2} = - \frac{4}{3} \alpha_0 (3 - c^2) \frac{1}{\frac{V_1}{V_0} + \phi_i^2} = - 0.00233\ee
As a result, the scalar to tensor ratio and the spectral index take the values $r=16\epsilon=0.0699$; and, $n_s = 1 - 6\epsilon + 2\eta = 0.969$ respectively, which are at par with the latest released data \cite{Pl1, Pl2}, while number of e-folds is sufficient to solve the horizon and flatness problem. Under the same choice of $c = 1$, and $\alpha_0 = \frac{M_{P}^2}{2}$, if we increase the magnitude of $\frac{V_1}{V_0}$, the value of $\mathrm{N}$ decreases, while, $r = 0.0699$ remains unaltered, and $0.969 \le n_s \le 0.972$ tally with the experimental data. These we tabulate underneath.

\begin{center} \textbf{Table - 1}\end{center}
\begin{center} Computed Inflationary parameters. Note that $r$ remains the same for a wide range of $\phi_i$.\end{center}

\begin{tabular}{|c|c|c|c|c|c|c|c|c|}
  \hline

  $\frac{V_1}{V_0}$ & $\phi_i$ &$\phi_f(\epsilon_f = 1)$& $\epsilon$ & $\eta$& $r$  & $n_s$ &~~$\mathrm{N}~~$ &$H_*$\\
  \hline
  $-$500 & 32.74 & 22.94 & 0.00437 & $-$0.00233 & 0.0699 & 0.969 & 71 & 1.69 $\times 10^{-5}$\\\hline
  $-$1000 & 41.54 & 32.20 & 0.00437 & $-$0.00184 & 0.0699 & 0.970 & 67 & 1.90 $\times 10^{-5}$\\\hline
  $-$2000 & 54.30 & 45.30  & 0.00437 & $-$0.00140 & 0.0699 & 0.971 & 64 & 2.18 $\times 10^{-5}$\\\hline
  $-$5000 & 79.98 & 71.29  & 0.00437 & $-$0.00095 & 0.0699 & 0.972 & 62 & 2.64 $\times 10^{-5}$\\\hline
  $-$10000 & 109.11  & 100.58 & 0.00437 & $-$0.00070 & 0.0699 & 0.972 & 60 & 3.09 $\times 10^{-5}$\\
  \hline
\end{tabular}\\
\\
\\
In table 1, $\frac{V_1}{V_0}$ is given in the unit of $M_P^2$, while, $\phi$ and the scale of inflation $H_*$ appear in the unit of $M_P$. $H_*$ has been computed in view of the first equation \eqref{slow004}, under the choice, $V_0 = 10^{-12} M_{P}^2$. Clearly the scale is sub-Planckian and matches with the energy scale for single field inflation \cite{SF},

\be H_* = 8 \times 10^{13}\sqrt{r\over 0.2} GeV = 4.733 \times 10^{13} GeV \approx 1.97 \times 10^{-5}M_P,\ee
taking into account $r = 0.0699$, as appeared in table 1. Now, as inflation ends, the scalar field should rapidly oscillate to produce particles. In the process, the universe transforms to the phase of `hot big-bang', containing a hot thick soup of plasma. This phenomena, dubbed as graceful exit, is needed for the structure formation along with the formation of CMB. Let us therefore check if the present model admits graceful exit from inflation. As inflation ends, $\dot\phi^2$ becomes comparable to $V(\phi)$ and therefore equation (\ref{slow004}) may be expressed as,

\be\label{slow005} \frac{1}{2}\dot\phi^2 + V = 2\alpha_0 (3 - c^2)H^2\ee
Now, choosing the same form of potential $V(\phi) = V_1 + V_0\phi^2$ and also the same numerical values for $c =1$, and  $\alpha_0 = \frac{M_{P}^2}{2}$, one can express the above equation (\ref{slow005}) as

\be\label{slow006} \frac{\dot\phi^2}{2V_1} + \Bigg(1 + \frac{V_0}{V_1}\phi^2\Bigg) = \frac{2H^2}{V_1}.\ee
Note that at the scale of inflation the numerical value of the factor $\left|{2 H_*^2\over V_1}\right| \approx 0.254$. At the end of inflation, as $\phi$ decreases, the Hubble rate $H$ decreases considerably, so that one can neglect the term on the right hand side, and approximate the above equation to,

\be\label{slow007} \dot\phi^2 + 2\Big(V_1 + V_0\phi^2\Big) = 0,\ee
which may be integrated to yield,

\be \phi = \frac{1}{2V_0}\Bigg(e^{i\sqrt{2V_0}~t} - V_0V_1e^{ - i\sqrt{2V_0}~t}\Bigg).\ee
Thus as inflation ends, the field starts oscillating many times over a Hubble time. Such coherent oscillating field corresponds to a condensate of non-relativistic massive (inflaton) particles, ensuring graceful exit from the inflationary regime. This drives a matter-dominated era at the end of inflation. Since $\sigma^2 = c^2 H^2 = H^2$ (for our choice, $c = 1)$, so the anisotropy is slowly varying too, but starts decreasing substantially to the extent of the Hubble parameter $H$, after inflation ends.\\

Although everything looks good with the above analysis, indeed there is a problem, which needs clarification. In the standard slow-roll inflation, observables can be computed by the background slow-roll parameters. However, the observables are not determined by the slow-roll parameters only, in generic inflationary scenarios. Such a typical case is the scenario in which perturbations have a non-trivial sound speed, since in that case, results depend on the sound speed. For example, the Weyl square term modifies the speed of gravitational waves. Further, in the multi-field case, the results depend on the trajectory and the isocurvature modes. Additionally, if the background is anisotropic, the observables should also depend on the anisotropy. It is therefore required to discuss the perturbations to compute the observables in the present case, as has been done earlier by several authors in different anisotropic cosmological models \cite{p1,p2,p3,p4,p5,p6,p7,p8}. However, it turns out to be extremely difficult to compute the same in the high-order theory under study. Hopefully, it may be done in the future.

\section{Concluding Remarks:}

The pre-inflationary stage of the universe could have been chaotic and anisotropic, which lead us to the present study. A novel attempt has been made to explore the evolution of early universe for higher order theory of gravity, starting from super-Planckian to the sub-Planckian era, relaxing the cosmological principle, by incorporating anisotropy in the background space-time. The two Hamiltonians obtained following modified Horowitz' formalism (which judiciously bypasses constraint analysis) and Dirac's constraint analysis (after taking care of the divergent terms appearing in the action) have been found to be identical. It is well-known that due to diffeomorphic invariance, time as well as probability interpretation (standard quantum mechanical) cease to exist in gravity. However, for the higher-order theories, an internal parameter (usually the proper volume) has been found to play the role of the time parameter, so that probability interpretation is straight forward, and the effective Hamiltonian operator turns out to be hermitian. In the present anisotropic model too, same results have been found, while the hermiticity of the effective Hamiltonian operator together with the continuity equation remove the arbitrariness of the operator ordering indices. This is important, since it was realized \cite{Stelle} that the presence of $R_{\mu\nu}R^{\mu\nu}$ term leads to ghosts when expanded in the perturbative series about the linearized theory, since the linearized energy of the five massive spin-2 excitations is negative definite, which destroys the unitarity. However, several non-perturbative analysis in different directions, revealed that such analysis might be naive and misleading \cite{Tomb77, Tomb80, Frad82, Tomb84, Anto86, Tomb15, Kaku83, Boul83}. We performed yet another non-perturbative analysis from the angle of quantum cosmology, and find that the effective Hamiltonian is hermitian. In this connection, we mention that if an operator is self-adjoint then only the dynamics is unitary and the time parameter can be extended to the real line (including negative values). However, a symmetric operator defined everywhere is necessarily bounded, i.e. a self-adjoint operator is hermitian, but not the reverse. Such a problem might appear in this regard here, since the volume plays the role of time parameter. Further, the operator $H_1$ of equation \eqref{Hsplit}, which is the momentum operator on the half-line ($y > 0$), perhaps is not self-adjoint \cite{Reed}. Hence, any wave packet moving along the axis $y$ would eventually reach the boundary $y = 0$ and start crossing it, leading to the disappearance of probability. This is a serious problem with the entire quantum framework because the evolution in that case is not unitary. However, we propose this non-unitary dynamics as a sort of approximation to quantum theory which is valid away from $y = 0$ and the proper volume, $a^3 = 0$. Further, if we consider that the theory has ghost degrees of freedom, then the present analysis suffers from unremovable pathologies. Note, one can change the variable, so that the Hamiltonian is linear in a momentum operator, which is a direct consequence of the Ostrogradski theorem. Indeed we split the Hamiltonian into a linear part and else. Thereafter we interpret the remaining part as the effective Hamiltonian in regard to the linear part. Thus, the ghost variable has been interpreted as the time parameter. The ghost part therefore, should be treated as a perturbation and splitting might not be justified. Further, in the perturbative analysis, the theory has been found to contain 5 ghost degrees of freedom due to massive spin-2 field. Hence, there might exist 5 time parameters as well, and the probability interpretation might not be justified. In that case, as we mentioned in the introduction, the present analysis is robust and may be performed in the case of ghost free bigravity model, as we have already applied it in the modified Lanczos-Lovelock gravity in isotropic and homogeneous background \cite{Ab9}.

In any case, the semiclassical wavefunction is strongly peaked around the classical de-Sitter solution, which reveals the emergence of classical trajectories, in accordance with Hartle' criterion. Further, sub-Planckian energy scale of inflation validates semiclassical approximation. The inflationary parameters show excellent agreement with the latest released Planck's data, and the theory admits graceful exit as well. This only proclaims the triumph and the simplicity of the inflationary scenario. However, several additional issues in connection with the present inflationary framework need to be addressed. In the isotropic and homogeneous background, the translational and rotational symmetries remain unbroken, while invariance under time translation is broken during inflation. In different anisotropic inflationary models, these symmetries are also broken. Some of these asymmetric models even predict a modification to the isotropic power spectra. Nevertheless, no evidence of dipolar asymmetry is evident, while temperature asymmetry has been confirmed to be due to a statistical fluctuation and not due to power law modulated power spectrum \cite{Pl2}. There is also no evidence of quadrupolar modulation too. In fact a slight negative value of $g_*$, which is a parameter characterizing the amplitude of violation of rotational symmetry, was detected \cite{Pl2}, but critical analysis revealed that this is solely due to the effect of WMAP’s asymmetric beams coupled with the scan pattern \cite{quad1, quad2}. In fact, after removing the effects of Planck’s asymmetric beams and Galactic foreground emission, no evidence for $g_*$ is found \cite{1310}. In a nutshell, there is no remnant of the pre-inflationary anisotropy, and no evidence of modified power spectrum (isotropic) is evident. In the present model, the measure of anisotropy for the space-time \eqref{anisonew} under consideration is given by the expression $\big|{\dot\chi\over \dot\xi}\big|$. Anisotropy is small provided $\big|{\dot\chi\over \dot\xi}\big|\ll 1$. In the solutions presented in \eqref{Sol2}, $\big|{\dot\chi\over \dot\xi}\big| = {c\over \sqrt 3}$, while we choose $c = 1$ to study inflation. Since, $\big|{\dot\chi\over \dot\xi}\big| < 1$, anisotropy remains small during inflation, but not sufficient enough, and therefore inflation is essentially anisotropic. It is important to mention that, $c$ cannot be made small enough, to ensure nearly isotropic inflation, since in that case, the number of e-folds becomes substantially large, and the universe becomes super-cold at the end of inflation. As already mentioned, indeed, it is required to discuss the perturbations to compute observables, since we are dealing with non-standard model. However, we find it extremely difficult to compute it for the model under consideration and therefore leave it at present. Nevertheless, it may be mentioned that in the anisotropic inflation under consideration, our present analysis is straightforward, and anisotropy does not play any vital role during inflation. Thus, there is supposed to be no deviation from results computed by perturbations, and statistical isotropy of the primordial structure of the universe, supposedly remains preserved. Although anisotropy is small in the present case, anisotropic expansion rate being proportional to the Hubble expansion rate ($\sigma = c H$) is also slowly varying, and so is not reduced much at the end of inflation. In fact, at the end of inflation, $\sigma = H_{end} \approx 4.2\times 10^{-6} M_p$, since we choose $c = 1$. However, it was shown long back in several anisotropic models with imperfect fluid sources (viscous and heat flux), that such trace of anisotropy is considerably reduced during the evolution of the universe in the matter dominated era \cite{AB1, AB2, AB3, AB4, AB5, AB6}.

\appendix

\section{Modified Horowitz' Formalism:}

As mentioned earlier in subsection (4.1), that the lapse function $N$ appears in the action \eqref{A2} with its time derivative, unlike GTR. This uncanny behaviour insists to treat the lapse function as a dynamical variable, although, it is essentially a gauge, and this issue desists from establishing diffeomorphic invariance. However, one can easily compute the Hessian determinant to be sure that it vanishes, making the action degenerate. Thus, canonical formulation requires to handle the situation following Dirac's algorithm of constraint analysis, which we performed in the main text. Indeed it is true, but as mentioned, there is an alternative, by the name MHF, in which one can bypass Dirac's programme, by introducing canonical auxiliary variables, and at the end switch over from the auxiliary variables to the basic variables $K_{ij}$. Although, equivalence between Dirac's and MHFs has been established in anisotropic models too \cite{Ani}, still for the sake of completeness, in the appendix, we show that the two are equivalent in the present case also. This makes one free to choose any of the two formalisms.\\

In MHF, as mentioned, the action is first expressed in terms of the basic variables $h_{ij}$, divergent terms are then removed upon integration by parts, which are cancelled with the supplementary boundary terms. Thus, we can initiate the programme with action \eqref{A2}, to find auxiliary variables, taking the derivative of the action \eqref{A2} with respect to the highest derivative present in it. In the present case the auxiliary variables $Q_1$, and $Q_2$ are,

\be\label{aux}\begin{split} &Q_1 = N {\partial A\over \partial \ddot z} = 3\frac{\sqrt{z}}{N^2}\Bigg[3\beta\Bigg(2\frac{\ddot z}{z} + \frac{\dot y^2}{y^2} - 2\frac{\dot N\dot z}{Nz}\Bigg) + \gamma\Bigg(2\frac{\ddot z}{z} + \frac{3\dot y^2}{2y^2} - 2\frac{\dot N\dot z}{Nz}\Bigg)\Bigg],\\
&Q_2 = N {\partial A\over \partial \ddot y} = 3\gamma\Bigg(\frac{\ddot y}{y} + \frac{3\dot z\dot y}{2zy} - \frac{\dot N\dot y}{Ny}\Bigg)\frac{{z}^{3\over2}}{yN^2}.\end{split}\ee
Now, substituting the auxiliary variables \eqref{aux} judiciously into the action (\ref{A2}) one obtains,

\be\label{AM1}\begin{split} A = \int\Bigg[\frac{Q_1\ddot z}{N}& + \frac{Q_2\ddot y}{N} - \frac{Q_1\dot N\dot z}{N^2} - \frac{Q_2\dot N\dot y}{N^2} - \frac{N\sqrt zQ_1^2}{12(3\beta + \gamma)} - \frac{Ny^2Q_2^2}{6\gamma z^{3\over2}} + \frac{3(2\beta + \gamma)zQ_1\dot y^2}{4(3\beta + \gamma)Ny^2} + \frac{3Q_2\dot y\dot z}{2Nz}\\
& - \Bigg\{\frac{3\gamma(12\beta + 5\gamma)\dot y^4}{16(3\beta + \gamma)N^2y^4} + \frac{3\alpha}{2}\Bigg(\frac{\dot z^2}{z^2} - \frac{\dot y^2}{y^2}\Bigg) + \frac{3\gamma}{N^2}\Bigg(\frac{3\dot y^2\dot z^2}{4y^2z^2} + \frac{\dot y^3\dot z}{y^3z}\Bigg) + 3\alpha'\dot\phi\frac{\dot z}{z} \\
& - \frac{\gamma'\dot\phi}{N^2}\Bigg(\frac{\dot z^3}{2z^3} + \frac{\dot y^3}{y^3} \Bigg) - \frac{\dot\phi^2}{2} + N^2V\Bigg\}\frac{{z}^{3\over2}}{N}\Bigg]dt.\end{split}\ee
The rest of the total derivative terms, viz., $\big(\frac{Q_1\dot z}{N} + \frac{Q_2\dot y}{N}\big)$ are integrated out by parts yet again, and action \eqref{AM1} is finally expressed as,

\be\label{AM2}\begin{split} A = \int&\Bigg[-\frac{\dot Q_1\dot z}{N} - \frac{\dot Q_2\dot y}{N} - \frac{N\sqrt zQ_1^2}{12(3\beta + \gamma)} - \frac{Ny^2Q_2^2}{6\gamma z^{3\over2}} + \frac{3(2\beta + \gamma)zQ_1\dot y^2}{4(3\beta + \gamma)Ny^2} + \frac{3Q_2\dot y\dot z}{2Nz} - \Bigg\{\frac{3\gamma(12\beta + 5\gamma)\dot y^4}{16(3\beta + \gamma)N^2y^4}\\
& + \frac{3\alpha}{2}\Bigg(\frac{\dot z^2}{z^2} - \frac{\dot y^2}{y^2}\Bigg) + \frac{3\gamma}{N^2}\Bigg(\frac{3\dot y^2\dot z^2}{4y^2z^2} + \frac{\dot y^3\dot z}{y^3z}\Bigg) + 3\alpha'\dot\phi\frac{\dot z}{z} - \frac{\gamma'\dot\phi}{N^2}\Bigg(\frac{\dot z^3}{2z^3} + \frac{\dot y^3}{y^3} \Bigg) - \frac{\dot\phi^2}{2} + N^2V\Bigg\}\frac{{z}^{3\over2}}{N}\Bigg]dt.\end{split}\ee
In the process $\dot N$ disappears from the action. However, the canonical momenta,

\be\label{momenta MHF}\begin{split}
& p_z =  - \frac{\dot Q_1}{N} + \frac{3Q_2\dot y}{2Nz} - \Bigg\{3\alpha\frac{\dot z}{z} + \frac{3\gamma}{N^2}\Bigg(\frac{3\dot y^2\dot z}{2y^2z} + \frac{\dot y^3}{y^3}\Bigg) + 3\alpha'\dot\phi - \frac{3\gamma'\dot z^2\dot\phi}{2N^2 z^2}\Bigg\}\frac{\sqrt z}{N};\\
& p_y = - \frac{\dot Q_2}{N} + \frac{3(2\beta + \gamma)zQ_1\dot y}{2(3\beta + \gamma)Ny^2} + \frac{3Q_2\dot z}{2Nz} - \Bigg\{\frac{3\gamma(12\beta + 5\gamma)\dot y^3}{4(3\beta + \gamma)N^2y^3} - 3\alpha\frac{\dot y}{y} + \frac{9\gamma}{N^2}\Bigg(\frac{\dot y\dot z^2}{2yz^2} + \frac{\dot y^2\dot z}{y^2z}\Bigg) - \frac{3\gamma'\dot y^2\dot\phi}{N^2 y^2}\Bigg\}\frac{{z}^{3\over2}}{yN};\\
& p_\phi = \Bigg\{\dot\phi - 3\alpha'\frac{\dot z}{z}  + \frac{\gamma'}{N^2}\Bigg(\frac{\dot z^3}{2z^3} + \frac{\dot y^3}{y^3} \Bigg)\Bigg\}\frac{z^{3\over2}}{N};\hspace{0.1in}p_{Q_1} = - \frac{\dot z}{N}; \hspace{0.1in}p_{Q_2} = - \frac{\dot y}{N}; \hspace{0.1in}p_N = 0,\end{split}\ee
clearly signal that the action is degenerate, since $\dot N$ cannot be expressed in terms of $P_N$. Nevertheless, the expression for the constraint Hamiltonian reads as:

\be\label{HM}\begin{split} H_c &= \dot zp_z + \dot Q_1p_{Q_1} + \dot yp_y + \dot Q_2p_{Q_2} + \dot \phi p_\phi + \dot Np_N  - L\\
&= -\frac{\dot Q_1\dot z}{N} - \frac{\dot Q_2\dot y}{N} + \frac{N\sqrt zQ_1^2}{12(3\beta + \gamma)} + \frac{Ny^2Q_2^2}{6\gamma z^{3\over2}} + \frac{3(2\beta + \gamma)zQ_1\dot y^2}{4(3\beta + \gamma)Ny^2} + \frac{3Q_2\dot y\dot z}{2Nz} - \Bigg\{\frac{9\gamma(12\beta + 5\gamma)\dot y^4}{16(3\beta + \gamma)N^2y^4}\\&~~~
+ \frac{3\alpha}{2}\Bigg(\frac{\dot z^2}{z^2} - \frac{\dot y^2}{y^2}\Bigg)
+ \frac{9\gamma}{N^2}\Bigg(\frac{3\dot y^2\dot z^2}{4y^2z^2} + \frac{\dot y^3\dot z}{y^3z} \Bigg) + 3\alpha'\dot\phi\frac{\dot z}{z} - \frac{3\gamma'\dot\phi}{N^2}\Bigg(\frac{\dot z^3}{2z^3} + \frac{\dot y^3}{y^3}\Bigg) + \frac{\dot\phi^2}{2} + N^2V\Bigg\}\frac{{z}^{3\over2}}{N},\end{split}\ee
which is also free from $\dot N$ and as a result Dirac's constraint analysis is bypassed, revealing the fact that the Hamiltonian should be cast in such a manner, so that $N$ finally appears as a Lagrange multiplier. It is now possible in principle, to translate the velocities in terms of momenta in view of \eqref{momenta MHF}, but that requires tedious algebra to perform. On the contrary, finding the expressions $p_{Q_1}p_z$ and $p_{Q_2}p_y$ in view of the definitions of momenta (\ref{momenta MHF}) and substituting the same into the constraint Hamiltonian (\ref{HM}), it is much easier to express the constraint Hamiltonian (\ref{HM}) in terms of the phase-space variables as,

\be\label{HM1}\begin{split} H_M = N&\mathcal{H_M} = N\Bigg[- p_{Q_1}p_z - p_{Q_2}p_y + \frac{N\sqrt zQ_1^2}{12(3\beta + \gamma)} + \frac{Ny^2Q_2^2}{6\gamma z^{3\over2}} - \frac{3(2\beta + \gamma)zQ_1p_{Q_2}^2}{4(3\beta + \gamma)y^2} - \frac{3Q_2p_{Q_1}p_{Q_2}}{2z} + \frac{p_\phi^2}{2z^{3\over2}} \\&
+ Up_\phi + \Bigg\{\frac{U^2}{2} + \frac{3\gamma(12\beta + 5\gamma)p_{Q_2}^4}{16(3\beta + \gamma)y^4} + \frac{3\alpha}{2}\Bigg(\frac{p_{Q_1}^2}{z^2} - \frac{p_{Q_2}^2}{y^2}\Bigg) + 3\gamma\Bigg(\frac{3p_{Q_1}^2p_{Q_2}^2}{4y^2z^2} + \frac{p_{Q_1}p_{Q_2}^3}{y^3z}\Bigg) + V\Bigg\}z^{3\over2}\Bigg],\end{split}\ee
where, $U = - 3\alpha'\frac{p_{Q_1}}{z} + \frac{\gamma'}{N^2}\Bigg(\frac{p_{Q_1}^3}{2z^3} + \frac{p_{Q_2}^3}{y^3} \Bigg),$ and in the process, diffeomorphic invariance is established. Nonetheless, the appearance of momenta ${P_Q}_1$, and ${P_Q}_2$ with fourth degrees, prevents the Hamiltonian from constructing a viable quantization scheme. Thus, we now express the Hamiltonian in terms of the other basic variables $K_{ij}$. This is possible under the replacements: $Q_1$ by $p_x$, $Q_2$ by $p_w$, $p_{Q_1}$ by $-x$ and $p_{Q_2}$ by $-w$. These indeed are canonical transformations, since $p_{Q_1} = - \frac{\dot z}{N} = -x$, and $Q_1 = N\frac{\partial A}{\partial\ddot z} = N\frac{\partial A}{\partial\dot x}\frac{\partial \dot x}{\partial\ddot z} = N\times p_x\times\frac{1}{N} = p_x$. Thus, $\{x,p_x\} = \frac{\partial x}{\partial Q_1}\frac{\partial p_x}{\partial p_{Q_1}} - \frac{\partial x}{\partial p_{Q_1}}\frac{\partial p_x}{\partial Q_1} = 0 - (-1)\times 1=1$. Similarly, $p_{Q_2} = - \frac{\dot y}{N} = -w$, and $Q_2 = N\frac{\partial A}{\partial\ddot y} = N\frac{\partial A}{\partial\dot w}\frac{\partial \dot w}{\partial\ddot z} = p_w$, and thus, $\{w,p_w\} =1$. As a result, $U = 3\alpha'\frac{x}{z} - \frac{\gamma'}{N^2}\Bigg(\frac{x^3}{2z^3} + \frac{w^3}{y^3} \Bigg)$. The Hamiltonian therefore finally expressed as,

\be\label{HM2}\begin{split} \mathcal{H_M} =& xp_z + wp_y + \frac{\sqrt zp_x^2}{12(3\beta + \gamma)} + \frac{y^2p_w^2}{6\gamma z^{3\over2}} - \frac{3(2\beta + \gamma)w^2zp_x}{4(3\beta + \gamma)y^2} - \frac{3wxp_w}{2z} + \frac{p_\phi^2}{2z^{3\over2}} + U p_\phi\\
& + \Bigg\{\frac{U^2}{2} + \frac{3\gamma(12\beta + 5\gamma)w^4}{16(3\beta + \gamma)y^4} + \frac{3\alpha}{2}\Bigg(\frac{x^2}{z^2} - \frac{w^2}{y^2}\Bigg) + 3\gamma\Bigg(\frac{3w^2x^2}{4y^2z^2} + \frac{w^3x}{y^3z} \Bigg) + V\Bigg\}z^{3\over2}.\end{split}\ee
This is the Hamiltonian found in \eqref{HD} following Dirac's algorithm, and therefore equivalence between the two formalisms is established.

\section{Hermiticity of the effective Hamiltonian:}

Again for the sake of completeness we compute hermiticity of the rest of the terms, viz. $\hat H_1, ~\hat H_4, ~\hat H_5, ~\hat H_6$ and $~\hat H_7$, which had not been handled in the main text of subsection (4.1). Let us consider the first expression of \eqref{Hsplit}, integrate it by parts, ignore the first term $\left({2i\hbar\omega\over 3x\sigma^{1\over 3}}\right)\Psi\Psi^*$ due to fall-of condition, to obtain,

\be\begin{split} \int (\hat H_1\Psi)^*\Psi dy = \frac{2i\hbar w}{3x\sigma^{1\over3}}\int\frac{\partial\Psi^*}{\partial y}\Psi dy &= -\frac{2i\hbar w}{3x\sigma^{1\over3}}\int\Psi^*\frac{\partial\Psi}{\partial y} dy = \int \Psi^*\hat H_1\Psi dy\end{split}\ee
Thus, $\hat H_1$ is hermitian.  Let us now consider $\hat H_4$ of \eqref{Hsplit},

\be\begin{split} \int (\hat H_4\Psi)^*\Psi dx = - i\hbar\frac{(2\beta + \gamma)w^2\sigma^{1\over3}}{4(3\beta + \gamma)y^2}\Bigg(\int\frac{2}{x}\frac{\partial\Psi^*}{\partial x}\Psi dx - \int \frac{1}{x^2}\Psi^*\Psi dx\Bigg).\end{split}\ee
Under integration by parts and dropping the integrated out terms due to fall-of condition, we obtain,

\be\begin{split} \int (\hat H_4\Psi)^*\Psi dx = i\hbar\frac{(2\beta + \gamma)w^2\sigma^{1\over3}}{4(3\beta + \gamma)y^2}\Bigg(\int\frac{2}{x}\frac{\partial\Psi}{\partial x}\Psi^* dx - \int \frac{1}{x^2}\Psi^*\Psi dx\Bigg) = \int \Psi^*\hat H_4\Psi dx.\end{split}\ee
Thus, $\hat H_4$ is also hermitian. For $\hat H_5$,

\be\begin{split} \int (\hat H_5\Psi)^*\Psi d w = - \frac{i\hbar}{2\sigma}\Bigg(\int2w\frac{\partial\Psi^*}{\partial w}\Psi dw + \int \Psi^*\Psi dw\Bigg)\end{split}\ee
again under integration by parts and dropping the integrated out terms due to fall-of condition, we obtain,

\be\begin{split} \int (\hat H_5\Psi)^*\Psi dw = \frac{i\hbar}{2\sigma}\Bigg(\int2w\frac{\partial\Psi}{\partial w}\Psi^* dw + \int \Psi^*\Psi dw\Bigg) = \int \Psi^*\hat H_5\Psi dw\end{split}.\ee
This reveals that $\hat H_5$ is hermitian. Now, since $\hat H_6$ is trivially hermitian, let us finally consider $\hat H_7$,

\be\begin{split} \int (\hat H_7\Psi)^*\Psi d\phi = \frac{2i\hbar}{3}\Bigg[3\frac{\alpha_0}{\sigma} - \frac{\gamma_0}{x\sigma^{1\over3}}\Bigg(\frac{x^3}{2\sigma^2} + \frac{w^3}{y^3}\Bigg)\Bigg]\Bigg(\int2\phi\frac{\partial\Psi^*}{\partial \phi}\Psi d\phi + \int\Psi^*\Psi d\phi\Bigg).\end{split}\ee
As before, under integration by parts and dropping the integrated out terms due to fall-of condition, we obtain,

\be\begin{split} \int (\hat H_7\Psi)^*\Psi d\phi = \frac{2i\hbar}{3}\Bigg[3\frac{\alpha_0}{\sigma} - \frac{\gamma_0}{x\sigma^{1\over3}}\Bigg(\frac{x^3}{2\sigma^2} + \frac{w^3}{y^3}\Bigg)\Bigg]\Bigg(\int2\phi\frac{\partial\Psi}{\partial \phi}\Psi^* d\phi + \int\Psi^*\Psi d\phi\Bigg) = \int \Psi^*\hat H_7\Psi d\phi.\end{split}\ee
This indicates $\hat H_7$ is also hermitian.

\end{document}